\input harvmac.tex
\input epsf
\input labeldefs.tmp
\noblackbox

\newcount\figno
 \figno=0
 \def\fig#1#2#3{
\par\begingroup\parindent=0pt\leftskip=1cm\rightskip=1cm\parindent=0pt
 \baselineskip=11pt
 \global\advance\figno by 1
 \midinsert
 \epsfxsize=#3
 \centerline{\epsfbox{#2}}
 \vskip 12pt
 {\bf Fig.\ \the\figno: } #1\par
 \endinsert\endgroup\par
 }
 \def\figlabel#1{\xdef#1{\the\figno}}

\newdimen\tableauside\tableauside=1.0ex
\newdimen\tableaurule\tableaurule=0.4pt
\newdimen\tableaustep
\def\phantomhrule#1{\hbox{\vbox to0pt{\hrule height\tableaurule width#1\vss}}}
\def\phantomvrule#1{\vbox{\hbox to0pt{\vrule width\tableaurule height#1\hss}}}
\def\sqr{\vbox{%
  \phantomhrule\tableaustep
  \hbox{\phantomvrule\tableaustep\kern\tableaustep\phantomvrule\tableaustep}%
  \hbox{\vbox{\phantomhrule\tableauside}\kern-\tableaurule}}}
\def\squares#1{\hbox{\count0=#1\noindent\loop\sqr
  \advance\count0 by-1 \ifnum\count0>0\repeat}}
\def\tableau#1{\vcenter{\offinterlineskip
  \tableaustep=\tableauside\advance\tableaustep by-\tableaurule
  \kern\normallineskip\hbox
    {\kern\normallineskip\vbox
      {\gettableau#1 0 }%
     \kern\normallineskip\kern\tableaurule}%
  \kern\normallineskip\kern\tableaurule}}
\def\gettableau#1 {\ifnum#1=0\let\next=\null\else
  \squares{#1}\let\next=\gettableau\fi\next}

\tableauside=1.0ex \tableaurule=0.4pt

\def\Kahler{K\"{a}hler }


\lref\OSV{OSV}
\newbox\tmpbox\setbox\tmpbox\hbox{\abstractfont }
\Title{\vbox{\baselineskip12pt \hbox{CALT-68-2622}
\hbox{NSF-KITP-06-118}
}}
{\vbox{\centerline{Quantum Entanglement of Baby Universes}}}
\vskip 0.2cm

\centerline{Mina Aganagic,$^1$ Takuya Okuda,$^2$ and Hirosi Ooguri$^3$}
\vskip 0.5cm
\centerline{$^1$\it University of California, Berkeley, CA 94720, USA}
\vskip 0.1cm
\centerline{$^2$\it Kavli Institute for Theoretical Physics, University of California,}
\centerline{\it Santa Barbara, CA 93106, USA}
\vskip 0.1cm
\centerline{$^3$\it California Institute of Technology, Pasadena,
CA 91125, USA}
\vskip 1.3cm

\noindent

We study quantum entanglements of baby universes which appear in
non-perturbative corrections to the OSV formula for
the entropy of extremal black holes in Type IIA string theory
compactified on the local Calabi-Yau manifold defined as
a rank 2 vector bundle over an arbitrary genus $G$ Riemann surface.
This generalizes the result for $G=1$ in hep-th/0504221.
Non-perturbative terms can be organized into a sum over
contributions from baby universes, and the total wave-function is
their coherent superposition in the third quantized Hilbert space.
We find that half of the universes preserve one set of supercharges
while the other half preserve a different set, making the total
universe stable but non-BPS. The parent universe generates baby universes
by brane/anti-brane pair creation, and baby universes are correlated
by conservation of non-normalizable D-brane charges under the process.
There are no other source of entanglement of baby universes, and
all possible states are superposed with the equal weight.

\bigskip\bigskip
\Date{December, 2006}

\def\Kahler{K\"{a}hler }

\lref\OoguriZV{
  H.~Ooguri, A.~Strominger and C.~Vafa,
  `Black hole attractors and the topological string,''
  Phys.\ Rev.\ D {\bf 70}, 106007 (2004)
  [arXiv:hep-th/0405146].
}

\lref\proofthree{
  C.~Beasley, D.~Gaiotto, M.~Guica, L.~Huang, A.~Strominger and X.~Yin,
  ``Why $Z_{BH} = |Z_{top}|^2$,''
  arXiv:hep-th/0608021.
}

\lref\proofone{
  D.~Gaiotto, A.~Strominger and X.~Yin,
  ``From AdS$_3$/CFT$_2$ to black holes/topological strings,''
  arXiv:hep-th/0602046.
}

\lref\prooftwo{
  R.~Dijkgraaf, C.~Vafa and E.~Verlinde,
  ``M-theory and a topological string duality,''
  arXiv:hep-th/0602087.
}

\lref\prooffour{
  J.~de Boer, M.~C.~N.~Cheng, R.~Dijkgraaf, J.~Manschot and E.~Verlinde,
  ``A farey tail for attractor black holes,''
  arXiv:hep-th/0608059.
}

\lref\proofzero{
  A.~Dabholkar, F.~Denef, G.~W.~Moore and B.~Pioline,
  ``Precision counting of small black holes,''
  JHEP {\bf 0510}, 096 (2005)
  [arXiv:hep-th/0507014].
}

\lref\prooffive{
F.~Denef, talks given in Beijing and Aspen;
G.~W.~Moore, talk given in Stony Brook (2006).}

\lref\torus{
  C.~Vafa,
  ``Two dimensional Yang-Mills, black holes and topological strings,''
  arXiv:hep-th/0406058.
}

\lref\grosstaylor{
  D.~J.~Gross and W.~Taylor,
  ``Two-dimensional QCD is a string theory,''
  Nucl.\ Phys.\ B {\bf 400}, 181 (1993)
  [arXiv:hep-th/9301068].
}

\lref\DGOV{
  R.~Dijkgraaf, R.~Gopakumar, H.~Ooguri and C.~Vafa,
  ``Baby universes in string theory,''
  Phys.\ Rev.\ D {\bf 73}, 066002 (2006)
  [arXiv:hep-th/0504221].
}

\lref\bcovone{
  M.~Bershadsky, S.~Cecotti, H.~Ooguri and C.~Vafa,
   ``Kodaira-Spencer theory of gravity and exact results for quantum string
  amplitudes,''
  Commun.\ Math.\ Phys.\  {\bf 165}, 311 (1994)
  [arXiv:hep-th/9309140].
}
\lref\bcovtwo{
  E.~Witten,
  ``Quantum background independence in string theory,''
  arXiv:hep-th/9306122.
}

\lref\bcovthree{
  R.~Dijkgraaf, E.~Verlinde and M.~Vonk,
  ``On the partition sum of the NS five-brane,''
  arXiv:hep-th/0205281.
}

\lref\bcovfour{
  E.~Verlinde,
  ``Attractors and the holomorphic anomaly,''
  arXiv:hep-th/0412139.
}
\lref\bcovfive{
  H.~Ooguri, C.~Vafa and E.~Verlinde,
  ``Hartle-Hawking wave-function for flux compactifications,''
  Lett.\ Math.\ Phys.\  {\bf 74}, 311 (2005)
  [arXiv:hep-th/0502211].
}

\lref\AOSV{
  M.~Aganagic, H.~Ooguri, N.~Saulina and C.~Vafa,
   ``Black holes, $q$-deformed 2d Yang-Mills, and non-perturbative topological
  strings,''
  Nucl.\ Phys.\ B {\bf 715}, 304 (2005)
  [arXiv:hep-th/0411280].
}

\lref\anv{
  M.~Aganagic, A.~Neitzke and C.~Vafa,
  ``BPS microstates and the open topological string wave function,''
  arXiv:hep-th/0504054.
}

\lref\BP{
  J.~Bryan and R.~Pandharipande,
  ``The local Gromov-Witten theory of curves,''
  arXiv:math.ag/0411037.
}

\lref\other{D.~Jafferis and J.~Marsano,
  arXiv:hep-th/0509004.
X.~Arsiwalla, R.~Boels, M.~Marino and A.~Sinkovics,
  Phys.\ Rev.\ D {\bf 73}, 026005 (2006)
  [arXiv:hep-th/0509002],
N.~Caporaso, M.~Cirafici, L.~Griguolo, S.~Pasquetti, D.~Seminara and R.~J.~Szabo,
  JHEP {\bf 0601}, 035 (2006)
  [arXiv:hep-th/0509041],
N.~Caporaso, M.~Cirafici, L.~Griguolo, S.~Pasquetti, D.~Seminara and R.~J.~Szabo,
  J.\ Phys.\ Conf.\ Ser.\  {\bf 33}, 13 (2006)
  [arXiv:hep-th/0512213],
H.~Kanno,
  Nucl.\ Phys.\ B {\bf 745}, 165 (2006)
  [arXiv:hep-th/0602179],
S.~de Haro, S.~Ramgoolam and A.~Torrielli,
  arXiv:hep-th/0603056,
N.~Caporaso, M.~Cirafici, L.~Griguolo, S.~Pasquetti, D.~Seminara and R.~J.~Szabo,
  arXiv:hep-th/0609129,
L.~Griguolo, D.~Seminara, R.~J.~Szabo and A.~Tanzini,
  arXiv:hep-th/0610155,
P.~Zhang,
  arXiv:hep-th/0611115,
R.~A.~Laamara, A.~Belhaj, L.~B.~Drissi and E.~H.~Saidi,
  arXiv:hep-th/0611289,
F.~Fucito, J.~F.~Morales and R.~Poghossian,
  arXiv:hep-th/0610154.
}
\lref\wb{
  E.~Witten,
  ``Quantum background independence in string theory,''
  arXiv:hep-th/9306122.
}
\lref\VCK{
  E.~Verlinde,
  ``Attractors and the holomorphic anomaly,''
  arXiv:hep-th/0412139.
}

\lref\DAC{
  R.~Dijkgraaf, E.~Verlinde and M.~Vonk,
  ``On the partition sum of the NS five-brane,''
  arXiv:hep-th/0205281.
}

\lref\ABV{
  M.~Aganagic, V.~Bouchard and A.~Klemm,
  ``Topological strings and (almost) modular forms,''
  arXiv:hep-th/0607100.
}
\lref\AJS{
  M.~Aganagic, D.~Jafferis and N.~Saulina,
   ``Branes, black holes and topological strings on toric Calabi-Yau
  manifolds,''
  arXiv:hep-th/0512245.
}

\lref\GBZ{
  M.~Gunaydin, A.~Neitzke and B.~Pioline,
  ``Topological wave functions and heat equations,''
  arXiv:hep-th/0607200.
}

\lref\MacDonald{
I. G. ~Macdonald, ``Symmetric functions and Hall polynomials,'' Clarendon
Press, 1995.
}

\newsec{Introduction}

What distinguishes string theory from any other approaches to quantum
gravity is, among others, the fact that one can estimate non-trivial
quantum gravity effects in controlled approximations. The OSV
conjecture \OoguriZV, for example, allows one to evaluate quantum
corrections to the Bekenstein-Hawking area-entropy relation for BPS
black holes in four dimensions to all orders of the expansion in
powers of spacetime curvatures. The conjecture identifies each term in
the expansion to the topological string partition function of a given
genus, which is manifestly finite and can be computed explicitly as a
function of geometry of the internal Calabi-Yau manifold. The
conjecture has been tested in \refs{\torus,\AOSV, \AJS, \proofzero},
and general proofs of the conjecture have been presented recently by
several groups
\refs{\proofone,\proofthree,\prooffour,\prooffive}.

In some cases, one can also evaluate the entropy exactly by mapping
its computation to solvable counting problems in the dual gauge
theories. The large $N$ expansion of the gauge theory results can
then be used to test the OSV conjecture to all orders in the string perturbation
expansion.  Even better, by identifying the difference of the OSV
formula and the exact gauge theory results, one can learn about
non-perturbative quantum gravity effects, $e.g.$ how to sum over
spacetime topologies. The purpose of this paper is to apply this idea
to specific examples, where computation can be done explicitly.
We find that only quantum entanglement among baby universes
is the one required by charge conservation, and otherwise all
possible states are coherently superposed with the equal weight
for a given number of baby universes.

The main ingredients of the OSV formula are topological string
partition function $Z^{top}$ of the Calabi-Yau manifold ${\cal M}$ and the
partition function
$Z_{BH}$ of black holes obtained by wrapping D-branes on cycles of ${\cal M}$.
The topological string partition function $Z^{top}$ is
expressed perturbatively as
\eqn\toppartion{ Z^{top}(t) = \exp\left[ \sum_{g=0}^\infty F_g(t)\; g_s^{2g-2}\right],}
where $F_g$ is the $g$-loop vacuum amplitude, and
$t$'s are the (flat) coordinates
of the Calabi-Yau moduli space of ${\cal M}$; they are coordinates on
the complexified K\"ahler moduli space for the A-model (in IIA) and the complex
structure moduli space for the B-model (in IIB).  $g_s$ is the
topological string coupling constant.
The conjecture relates $Z^{top}$ to the Laplace transform of
the Witten index $\Omega(p,q)$ for supersymmetric ground states
of the extremal black hole with magnetic and electric
charges $(p,q)$ realized by wrapping D branes on cycles in ${\cal M}$.
More precisely, consider the black hole partition function
$$
Z_{BH}(p,\phi) = \sum_q \Omega(p,q) e^{-q\phi}
$$
where we fix the magnetic charges and sum over the electric ones.
The conjecture states
\eqn\osv{ Z_{BH}(p,\phi) = |Z^{top}(t, g_s)|^2. }
where the parameters of the two sides are
related by attractor mechanism as:
\eqn\attractor{g_s = {4 \pi i \over p^0+i\phi^0/\pi}, \qquad
t^i=  {1\over 2}\,( p^i + i \phi^i/\pi)\, g_s.}
In the above $p^{0}$ and $p^i$ are the magnetic charges
and $i$ runs over the number of moduli of the Calabi-Yau.
In IIA, for example, $p^0$ is the number of
D6 branes, and $p^i$ are the D4 brane charges.
The conjecture is supposed to hold to all orders in the string loop
expansion. However, the relation \osv\ may be corrected by effects
that are non-perturbative in the string coupling $g_s$.

The first concrete example of this was given in
\torus, where type IIA string theory on the Calabi-Yau manifold
${\cal M}$ which is a rank 2 vector bundle (in fact a sum of two line
bundles) over an
elliptic curve $\Sigma$ was studied. With $N$ D4 branes and no D6
branes on ${\cal M}$, the black hole partition function
$Z_{BH}$ turns out to be equal to the partition function of the 2d
Yang-Mills theory on the elliptic curve, which can be computed
exactly. The $g_s$ expansion of $Z_{BH}$ is identified with the
large $N$ expansion in the 2d Yang-Mills theory. It was shown
that, to all order in the $1/N$ expansion, the 2d Yang-Mills
partition function factorizes as
\eqn\twodym{ Z_{BH}(N, \phi) = \sum_{l\in  {\bf Z}}
    Z^{top}(t+ l g_s) \bar Z^{top}({\bar t} - l g_s),}
where $t = {1\over 2}( N + i \phi/\pi) g_s$.
The sum over $l$ is interpreted as
a RR flux through the elliptic curve on the base.
In particular,
one can identify the topological A-model string
theory on ${\cal M}$ as the large $N$ dual of the 2d Yang-Mills theory
studied earlier in \grosstaylor.

In this case, non-perturbative corrections to \twodym\
were evaluated in \DGOV. Taking into account terms non-perturbative
in $g_s$, one finds
\eqn\dgovexpansion{\eqalign{Z_{BH}(N,\phi) =
 & \sum_{k=1}^\infty (-1)^{k-1} C_k
\cr
& \qquad \sum_{l_1,\ldots l_k \in {\bf Z}}
Z^{top}(t_1+l_1 g_s)\bar Z^{top}({\bar t_1}-l_1g_s)
\cdots Z^{top}(t_k+l_k g_s)
 \bar Z^{top} ({
\bar t}_k - l_k g_s ),}}
where $C_k$ is the Catalan number given by
\eqn\catalan{
C_k = {(2k)! \over k! (k+1)!}
}
The $k=1$ term corresponds to the OSV formula \twodym , and the $k>1$
terms are interpreted as having $k$ disjoint baby universes. The
Catalan number simply counts the number of ways the baby universe can
be produced.  In the above,
$$
t_i = {1\over 2} \, (N_i + i \phi /\pi)g_s
$$
and for each $k$ baby universe configuration, the total D4 brane charge is $N$:
$$
\sum_{i=1}^{k} N_k = N.
$$

The topological string partition function $Z^{top}$ has an
interpretation as a wave-function in the Hilbert space ${\cal H}$
obtained by quantizing $H^3({\cal M})$ (in the mirror
 B-model language), as was explained in
\refs{\bcovone,\bcovtwo} and studied more recently in
\refs{\bcovthree,\ABV,\GBZ}. In particular, we can invert the OSV
formula \osv\ to write,
\eqn\inverseosv{ \Omega(p,q) = \int \,d\phi \;e^{q\phi} \;Z^{top}(p+i\phi/\pi)\,
\bar Z^{top}(p-i\phi/\pi),}
and regard the right-hand side as computing the norm-squared of
the wave-function \refs{\bcovthree,\bcovfour,\bcovfive},
$$\psi_{p,q}(\phi)=e^{q\phi/2}\;Z^{top}(p+i\phi/\pi).$$
$\psi_{p,q}(\phi)$ was interpreted in \refs{\bcovfive} as a
Hartle-Hawking type wave function for the universe defined on
a ${\cal M}\times S^2\times S^1$ spatial slice.
Generalizing this to the baby universe expansion,
\dgovexpansion\ was interpreted in \DGOV\
as a norm-squared of a wave-function in the third quantized Hilbert space
$\oplus_{k=1}^\infty {\cal H}^{\otimes k}$
given as a coherent superposition of baby universes.
The D-brane charges are
distributed among the baby universes in such a way that
the total charges are conserved.
Otherwise, no entanglement among the universes is found
in this case,
and all the states allowed by the charge conservation are summed
with the same weight within a sector of a given number of universes.

The purpose of this paper is to generalize the result of \DGOV\
for a case when ${\cal M}$ is the sum of two line bundles over
a genus $G$ Riemann surface $\Sigma_G$ for any $G$.
We mostly focus on $G>1$ since the $G=0$ is similar and we do not expect any new novelty.
In this case, the black
hole partition function $Z_{BH}$ is related to the partition function
of a certain deformation of the 2d Yang-Mills theory on $\Sigma_G$.
The precise deformation will be specified later. In \AOSV , the partition function $Z_{BH}$
is evaluated explicitly, and it was shown that the large $N$ expansion
of $Z_{BH}$ reproduces the OSV formula with one important subtlety.
The large $N$ factorization of $Z_{BH}$ produces topological string
partition function on ${\cal M}$, but with insertions of
additional ``ghost'' D-branes.
The ghost branes
generate correlations
between $\bar Z^{top}$ and $ Z^{top}$,
$i.e.$ the bra and ket vectors. Subsequently,
it was shown in \anv\ that the insertions of ghost D branes have a
dual $closed$ string interpretation. Namely,
because of the non-compactness of the Calabi-Yau manifold ${\cal M}$,
there are infinitely many K\"ahler moduli that are not supported by compact
2-cycles. We have to treat them at the same footing as the ordinary \Kahler moduli, as the
wave function $Z^{top}$ naturally depends on them.
Applying the OSV prescription to these moduli, with the corresponding electric and magnetic charges set to zero, generates the
correlation between $\bar Z^{top}$ and $ Z^{top}$.

In this paper, we will compute non-perturbative corrections to the
OSV formula for this geometry. As in the case of $G=1$ studied in
\DGOV, the black hole partition function $Z_{BH}$ is expressed as
a sum over Young diagrams with a fixed number of rows, and this
structure naturally leads to the holomorphic factorization \osv\ in
the limit when the number of rows is infinite. Non-perturbative
corrections arise due to the fact that the number of rows are in fact finite. We find that the non-perturbative terms can be organized
into a sum over baby universes as in \dgovexpansion\ with new
features. We find that half of baby universes preserve one set
of supercharges and the other half preserve a different set.
As a consequence, the total universe is not BPS, but it is still
stable since different baby universes are spatially disjoint.
Another feature is presence of a new type of ghost D-branes.
In addition to insertions of D-branes that
correlate the bra and ket vectors, as we mentioned at the end
of the previous paragraph, we
find yet another set of D-branes that correlate ket vectors
among themselves
(and another set of D-branes for bra vectors).
It turns out that these correlations also have their origin in
a conserved charge: namely the charge of the non-compact D2 branes
wrapping the fibers over the Riemann surface. The charges of these branes in all the universes have to add up to zero,
just as in the parent universe, and this induces correlations.
In the $G=1$ case studied in \DGOV, these branes and correlations associated to them were absent, as they would violate the toroidal symmetry of the base Riemann surface.
Moreover we find no other source of correlations
between different baby universes, for any $G$.

This paper is organized as follows. In section 2, we will summarize
our result. In section 3, we will present
expressions for $Z_{BH}$ and $Z^{top}$ for the rank 2 bundle over $\Sigma_G$
computed in \AOSV\ and \BP\ respectively, and we will review the large
$N$ factorization of $Z_{BH}$ discussed in \AOSV . In section 4, we will
identify non-perturbative corrections to $|Z^{top}|^2$
by taking into account the
finite size effects of Young diagrams. Various technical computations are relegated to the appendices.

\newsec{The main result}

Consider type IIA superstring theory compactified on a Calabi-Yau
manifold ${\cal M}$ that contains a compact 2-cycle that can shrink,
a Riemann surface $\Sigma$ of genus $G$. The local region near $\Sigma$ is a
non-compact Calabi-Yau manifold which is the total space of the sum of
two line bundles over $\Sigma$
$$
{\cal
L}_{-p}\oplus {\cal L}_{p+2G-2}\rightarrow \Sigma,
$$
for some $positive$ integer $p$.
Now wrap $N$ D4-branes on the divisor
$$
{\cal D}={\cal L}_{-p}\rightarrow \Sigma.
$$
The D4 branes can form BPS bound states
with $Q_2$ D2 branes wrapping $\Sigma$ and $Q_0$ D0 branes. In \AOSV,
the Witten index
$
\Omega(N, Q_2,Q_0) = \tr' (-1)^F
$
of the BPS bound states of these objects is computed by the
supersymmetric path integral of the D4 brane theory on ${\cal
D}$. (The $\tr'$ in the index refers to the fact that the zero modes
corresponding to center of mass motion of the D4 branes are removed in
the trace.) The black-hole partition function $Z_{BH}$ is then defined
by
$$
Z_{BH} = \sum_{Q_2, Q_0} \Omega(N, Q_2, Q_0) e^{-Q_0 \phi^0 - Q_2 \phi^1}.
$$
The large $N$ expansion of $Z_{BH}$ is evaluated with the following result:
\eqn\anovfactor{
\eqalign{ Z_{BH}(N,\phi^0,\phi^1) = & \sum_{l=-\infty}^\infty \int dU_1 \cdots dU_{|2G-2|}
Z^{top}(t+lpg_s ; U_1, \cdots, U_{|2G-2|})\cr
&~~~~~~~~~~~~\times \bar Z^{top}(\bar t-lpg_s ; U_1, \cdots, U_{|2G-2|}),
}}
where $Z^{top}$ in the right-hand side is the topological string
partition function with insertions of indefinite number of ``ghost''
D- branes
along $|2G-2|$ Lagrangian submanifolds of ${\cal M}$, each of which
intersects with ${\cal D}$ around $S^1$ in the fiber of ${\cal L}_{p+2G-2}$
over one of $|2G-2|$ points on $\Sigma$,
and $U_1, \cdots, U_{|2G-2|}$ are holonomies taking value in $U(\infty)$.
The sum over $l$ is interpreted as a sum over 2-form flux through $\Sigma$.
The topological string coupling $g_s$ and the K\"ahler moduli $t$
associated to the size of $\Sigma$ are given by
$$
g_s = {4\pi i \over \phi^0}, ~~t = {1\over 2}
(p+2G-2) N g_s+ i {\phi^1 \over 2\pi}g_s.
$$
Note that the effective D4 brane charge is larger by
a factor of $p+2G-2$. The  magnetic
charge of a D4 brane on a divisor ${\cal D}$ is set by the intersection number
of ${\cal D}$ with ${\Sigma}$, the two-cycle wrapped by the
D2-branes. In the present context, the intersection number
is
$$
{\#({\cal D} {\cap} \Sigma)} = p+2G-2,
$$
and hence the above.

How is the formula \anovfactor , with the integral over $U$'s,
compatible with the OSV conjecture?
The Calabi-Yau manifold ${\cal M}$ is non-compact, and there are
infinitely many K\"ahler moduli which are not supported by compact
2-cycles.  According to \anv, eigenvalues of $U_1, ..., U_{|2G-2|} \in
U(\infty)$, are interpreted as exponentials of these K\"ahler
moduli.
The imaginary parts of the K\"ahler moduli are chemical potentials for
the electric charges and the integral over $U$'s sets all the electric
charges to be zero. The only nonzero magnetic charges are those
induced by the D4 branes and the attractor mechanism shifts the real
parts on the K\"ahler moduli (rescaling $U$'s).  The magnetic charge of the D4 branes
under a global $U(1)$ corresponding to a non-normalizable modulus is
given by their intersection number with the corresponding non-compact
2-cycle, just as in the compact 2-cycle case.  This turns out to be
non-zero here, giving the K\"ahler
moduli a non-vanishing real part.  The
integral over the holonomies should thus be thought of as an integral
over these non-normalizable K\"ahler moduli. It ensures that the gravity
computation in the right-hand side of \anovfactor\ matches with the
gauge theory computation of $Z_{BH}$ where we only count bound states
of D4 branes on ${\cal D}$ with D2 branes wrapping $\Sigma$ and D0
branes on $\Sigma$.

In this paper, we will follow the prescription of \DGOV\
to identify corrections to \anovfactor\ that are non-perturbative
in the topological string coupling $g_s$.
As we will show in subsection 4.2, the first correction to \anovfactor\ comes from configurations where
the parent universe splits into two by creating a baby universe.  The
parent universe still ends up carrying charges corresponding to $N$ D4
branes on
${\cal D}$
which it had
originally. But now,
it also carries charges of $-k$ branes on
$$
{\cal D}' = {\cal L}_{p+2G-2} \rightarrow \Sigma,
$$
with the baby universe carrying of the other $k$
units of this charge.
Let us denote by $(m,m')$
the charge of a universe corresponding to $m$ D4 branes on ${\cal D}$ and $m'$ D4 branes on ${\cal D}'$.
Then, this process corresponds to
$$
(N,0) \rightarrow (N,-k) \oplus (0,k).
$$
Here, $k$ is positive, as is needed\foot{In particular, note that the charges
of the branes need to be of the same sign for the branes to
be mutually BPS. Only in that case does the electrostatic repulsion cancel the gravitational attraction of the particles, so the net force condition is satisfied.}
in order for the first universe to correspond to a mutually BPS configuration
of D4 branes. This is because
the effective D4 brane charge
for branes wrapping ${\cal D}'$ is negative:
$$
{\#\,({\cal D}'\, {\cap}\, \Sigma)} = -p.
$$
Note that, even though this looks like we have pair created
$k$ branes and
anti-branes on ${\cal D}'$, this is true only at the level of the charges. In particular, all these universes are dual to
a single theory of $N$ D4 branes on ${\cal D}$.
The next leading correction comes from configurations
with 3 universes, where the second universes splits of a baby
$$
(N,0) \rightarrow (N,-k) \oplus (0,k)\rightarrow
(N,-k) \oplus (-m,k) \oplus (m,0)
$$
while creating $m$ units of D4 brane and anti-D4 brane charge on
${\cal D}$.
This will be demonstrated in subsection 4.4.
The configuration of branes in the
second universe is now supersymmetric provided
$m\geq 0$.
This also gives a natural ordering of the universes so that
odd (even) universes carry positive (negative) D4 brane charges.
This in particular implies that the odd and even universes
preserve different sets of supercharges.  If odd universes are BPS,
even universes are anti-BPS.  The total universe is still stable since
baby universes are disjoint.  One consequence of this is that a ket
vector for an even universe is $\bar Z^{top}$ while a ket vector
for an odd universe is $Z^{top}$.  This follows from the derivation
of the OSV conjecture in \proofone\ applied to the two types of baby
universes.  It also follows from the fact that the CPT conjugation is
an anti-unitary operation, and it exchanges bra and ket vectors.

In addition to this general structure, there are two more important
features.  As we mentioned in the above, for a single universe studied
in \AOSV , \anovfactor\ is imposing the
constraint of vanishing electric charges for the
non-normalizable \Kahler moduli, and this correlates
the bra and ket
vectors ($Z^{top}$ and $\bar Z^{top}$).
With more than one universe, this
structure is duplicated, setting the net electric charges to
zero. This induces correlations between the ket and the bra vectors of
the different universes.

Furthermore, for configurations with multiple universes,
another set of ghost D-branes appears that introduce correlations
between ket vectors (and another copy of this for the bra vectors).
In the following sections, we will show that these are also a
consequence of charge conservation -- this time for charges of the
non-compact D2 branes that can form bound states with the D4 branes. These
are also a consequence of the non-compactness of the manifold,
because they correspond to having different boundary conditions at
infinity of the Calabi-Yau ${\cal M}$.

\newsec{$q$-deformed Yang-Mills, large $N$ factorization, and topological string}

In this section we will review the results of \AOSV\ computing the
large $N$ limit of partition function $Z_{BH}$ of D-branes on ${\cal
M}$, and relation of this to topological strings on ${\cal M}$\foot{For related recent work see \other .}.

\subsec{Superstring and $q$-deformed Yang-Mills}

Consider type IIA superstring theory compactified on a Calabi-Yau
manifold ${\cal M}$ as in the previous section,
%
%
where we wrap $N$ D4-branes on the divisor ${\cal D}={\cal
L}_{-p}\rightarrow \Sigma$. The D4 branes can form BPS bound states
with $Q_2$ D2 branes wrapping $\Sigma$ and $Q_0$ D0 branes. The
indexed degeneracies
$
\Omega(N, Q_2,Q_0)
$
of the BPS bound states of these objects can be computed by the supersymmetric
path integral of the D4 brane theory on ${\cal D}$, in a topological sector with
$$
Q_0 = {1\over 8 \pi^2} \int_{\cal D} {\rm Tr} F \wedge F, \qquad Q_2 =
{1\over 2 \pi} \int {\rm Tr} F \wedge k .
$$
where $k$
is the normalized K\"{a}hler form such that $\int_\Sigma k=1$.  Computing the path integral without fixing the topological sector
thus automatically sums over all the D2 and the D0 brane charges.
To keep track of them, we will add to the action the terms
$$
{1\over 2g_s}\int{\rm tr}F\wedge
F+{\theta\over g_s}\int {\rm tr}F\wedge k.
$$
Then, ${4\pi^2 \over
g_s}$ and ${2\pi \theta\over g_s}$ are the chemical potentials for D0
and the D2 brane charges. We called these $\phi^0$, $\phi^1$ in section 1.

Since ${\cal D}$ is curved the supersymmetric theory on it automatically topologically twisted,
and it corresponds to the Vafa-Witten twist of the ${\cal N}=4$ theory in ${\cal D}$.
As argued in \refs{\torus, \AOSV}, since the manifold is non-compact and has $U(1)$ isometry
corresponding to rotating the fiber, the twisted ${\cal N}=4$ theory is in turn
localizes to a gauge theory on the Riemann surface $\Sigma$. This theory is
a certain deformation of the ordinary bosonic 2d $U(N)$ Yang-Mills theory,
a ``quantum'' Yang-Mills theory (qYM) on $\Sigma$.
This
is described in detail in \AOSV\ and we will only quote the results here.
%
In particular, the difference does not spoil the solvability of the two-dimensional bosonic Yang-Mills: the sewing and gluing prescription of the amplitudes still applies.
The qYM partition function on any Riemann surface can be obtained
by gluing the cap ${\cal C}$ and the pant $\cal P$ amplitudes
$$
Z({\cal C})_{\cal R} = S_{0{\cal R}}(g_s,N),
$$
$$
Z({\cal P})_{\cal R} = 1 / S_{0{\cal R}}(g_s,N)
$$
where $U(N)$ representations ${\cal R}$ correspond to the states in the Hilbert space of the 2d theory,
and
$$
S_{0{\cal R}} =\prod_{1\leq i<j\leq N}{[{\cal R}_i-{\cal R}_j+j-i]}.
$$
Here $[n]$ denotes the quantum number $n$, $[n] = q^{n\over 2}-q^{-{n\over 2}}$
with $q=e^{-g_s}$.
Note that $S_{0{\cal Q}}$ is similar to a $U(N)$ WZW S-matrix element,
but with non-integer level.

The full partition function of the modified Yang-Mills on the Riemann surface $\Sigma$
of genus $G$ is
\eqn\divy{
Z^{qYM}(\Sigma)= z^{YM} \sum_{\cal R}
\left(S_{0{\cal R}}\right)^{2G-2}
q^{{p\over 2}C_2({\cal R})} e^{i\theta C_1({\cal R})},
}
where
$$
C_2({\cal R})=\sum_{i=1}^N {\cal R}_i ({\cal R}_i-2i+N+1),~~C_1({\cal R})=\sum_{i=1}^N {\cal R}_i,
$$
are the $U(N)$ casimirs of the representation ${\cal R}$.
The precise normalization factor $z^{YM}$ will be discussed later.
Note that
$$
S_{0{\cal R}}/S_{00} = \dim_{q}({\cal R})
$$
is a quantum deformation of the dimension of a $U(N)$ representation ${\cal R}$.
The theory here differs from the ordinary two-dimensional Yang-Mills
in that the ordinary dimension of representation is replaced by its quantum deformation (for a review of ordinary 2d YM, see for example
\ref\moore{
S.~Cordes, G.~W.~Moore and S.~Ramgoolam,
``Lectures On 2-D Yang-Mills Theory, Equivariant Cohomology And Topological
Field Theories,''
Nucl.\ Phys.\ Proc.\ Suppl.\  {\bf 41}, 184 (1995)
[arXiv:hep-th/9411210].
}).
Because of this the theory at hand was termed the $q$-deformed
Yang-Mills theory in \AOSV . When the Riemann surface is a torus, as in \refs{\torus, \DGOV}
the difference goes away.

In the next subsection we will
derive the large $N$ limit of quantum YM amplitudes
from their description in terms of free fermions.

\subsec{Fermions and qYM Amplitudes}

The qYM partition function on any Riemann surface can be obtained
by gluing the cap ${\cal C}$ and the pant $\cal P$ amplitudes
$$
Z({\cal C})_{\cal R} = S_{0{\cal R}}(g_s,N),
$$
$$
Z({\cal P})_{\cal R} = 1 / S_{0{\cal R}}(g_s,N)
$$
Knowing the large $N$ limit of these two amplitudes, we can
compute the large $N$ limit of the YM amplitude on any Riemann
surface. This was done in \AOSV . In the present context, we
need to generalize their results to compute non-perturbative,
multi-center black hole corrections. To do that, a different derivation is more
adept.

There is a well known correspondence between $U(N)$ representations ${\cal R}$
and states in the Hilbert space of ${N}$ free, non-relativistic  fermions on a circle
(for a review, see for example \moore ). The representation ${\cal R}$ captures the momenta of the $N$ fermions
in the following way. If ${\cal R}_i$, $i=1,\ldots N$ are the lengths of the
rows of the $U(N)$ Young-tableaux corresponding to ${\cal R}$, than the $i$-th fermion
momentum is
$$
n_i = {\cal R}_i -i +{N+ 1 \over 2}.
$$
Recall that $U(N)$ differs from $SU(N)$ in that ${\cal R}_i$ are
not necessarily positive. The degree of freedom corresponding to
shifting all the fermion momenta and all ${\cal R}_i$'s by a
constant, corresponds to the choice of the $U(1)$ charge of the
representation. Recall that any $U(N)$ representation ${\cal R}$
can be obtained by tensoring an $SU(N)$ representation $R$ by
${l}$ powers of the determinant representation, and
correspondingly, we have
$$
{\cal R}_i = R_i +l.
$$
The $U(1)$ charge $q$ of the representation is then $q = N l+ |R|$.

For this fermion state, the explicit form of the master amplitude
is
\eqn\fund{
S_{O{\cal R}} = \prod_{1\leq i < j \leq N} {[{\cal R}_i - {\cal R}_j+j-i]}
}
Note that this is independent of the $U(1)$ charge.

\subsec{Large $N$ factorization}

At large $N$ the fermion states are described by independent fluctuations
about two well separated fermi surfaces.
Correspondingly, the representation
${R}$ can be thought of as a coupled representation
${R^+} \bar{R}^-$, where $R^+$ and $R^-$ describe the fermion
excitations about the two fermi surfaces.
\bigskip
\centerline{\epsfxsize 3.0truein\epsfbox{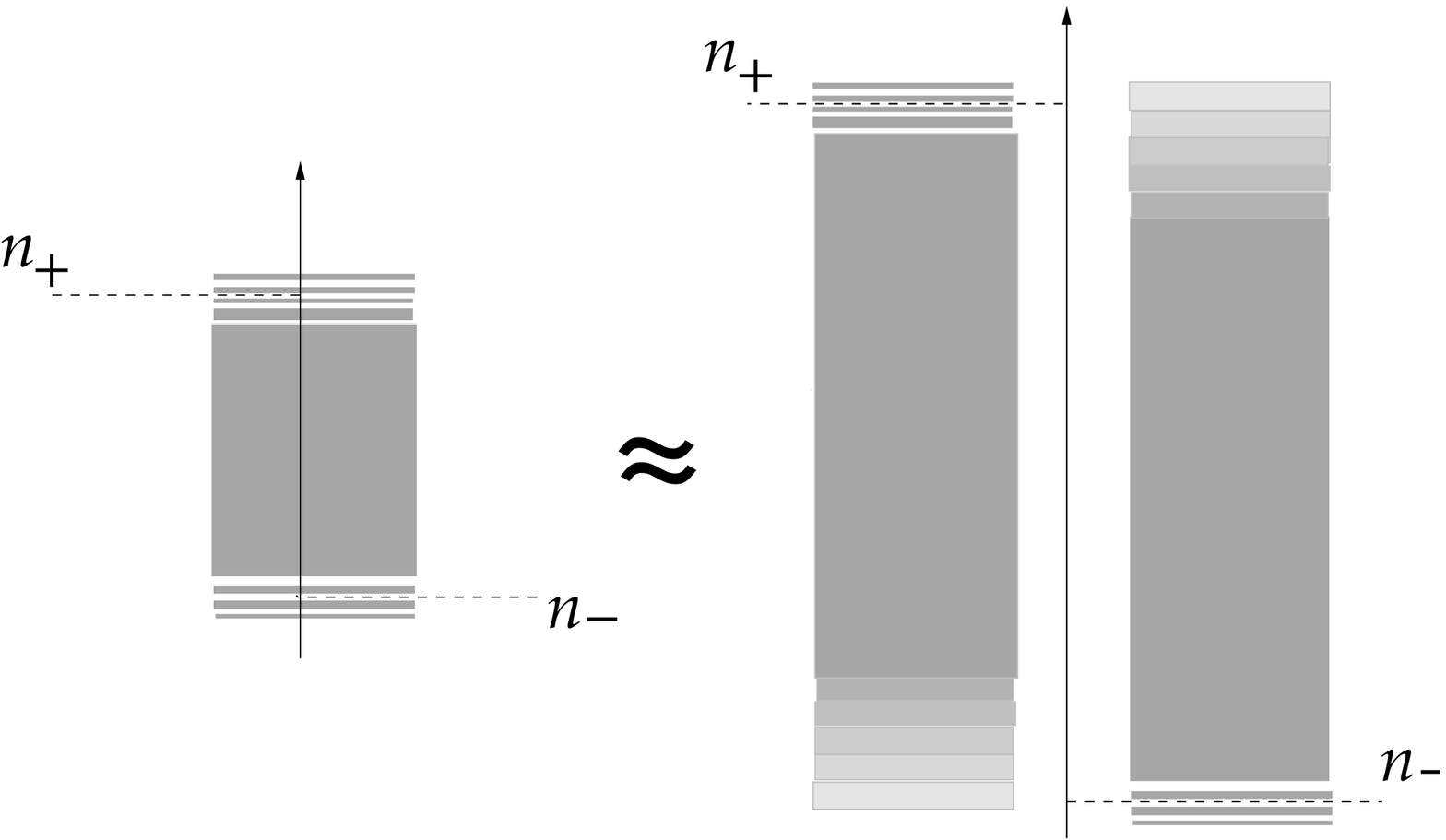}}
 \noindent{\ninepoint\sl
\baselineskip=2pt {\bf Fig.1}
{{Excitation about the top and the bottom fermi sea are
independent at large $N$, although they are not uncorrelated.
They are described by topological and anti-topological string amplitudes.
}}}
\bigskip

The amplitude $S_{0,{\cal R}}/S_{0,0}$ has pairwise interactions
of $excited$ fermions. Thus, the $N$ dependence of the amplitude
should be localized to the interactions of the fermions at the top
and bottom of the fermi seas only. So, we can guess that the
amplitude factorizes at large $N$ as:
$$\eqalign{
S_{O{\cal R}}/S_{00}(N,g_s)& =
\prod_{1\leq i < j \leq \infty} {[R^+_i - {R^+_j}+j-i]\over [j-i]}
\prod_{1\leq i < j \leq \infty}{[{R^-_i} - {R^-_j}+j-i] \over [j-i]}\cr
&\times \prod_{1\leq i ,j \leq \infty}
{[{R^+_i} + {R^-_j}+ {n^+-n^-} -i-j+1]\over [{n^+} - {n^-} -i-j+1]}}.
$$
All the interactions depend only on the distances,
in momentum space, of the between the excited fermions.
We have denoted by $n^+$ and $n^-$ the positions of the fermi surfaces (see Fig. 1), where
$$
{n^+}-{n^-}=N.
$$
We will prove this formula in appendix A.

We can now easily reproduce the results of \refs{\AOSV}. First, note that
the first and the second factors are given in terms of
Schur functions $s_R(q)$ (see appendix A. for a derivation)\foot{
The Schur functions $s_R$ are simply extensions of $U(N)$
characters to infinite rank. We will need only rudimentary properties of these functions. For more details about properties of Schur functions,
see \refs{\MacDonald} .}, which in turn correspond to topological string amplitudes:
$$
W_{0,R}= s_R(q^{\rho})=
 q^{k_R/4} \prod_{1\leq i < j \leq \infty} {[R_i - {R}_j+j-i]\over [j-i]}  \;
$$
where
$$
k_R =2 \sum_{\tableau{1} \in R} (j(\tableau{1})-i(\tableau{1})),
$$
as $i$ runs over rows, and $j$ over columns.
Here,
$$
q^{\rho} = (q^{-1/2},q^{-3/2},\ldots)
$$
One basic property of Schur functions is
that
\eqn\dtup{
\prod_{i,j}(1-x_i y_j) = \sum_{Q} (-)^{|Q|} s_Q(x) s_{Q^T}(y).
}
and
\eqn\dtdn{
\prod_{i,j}{(1-x_i y_j)^{-1}} = \sum_{Q} s_Q(x) s_{Q}(y).
}
Using \dtup , we can write the last factor above as
$$
(-1)^{|R^+| + |R^-|}
q^{-{1\over 2} N(|R^+|+|R^-|)} q^{-{1\over 4} (k_{R^+}+k_{R^-})}
\sum_{Q}(-)^{|Q|} s_Q(q^{\rho + R^+}) q^{|Q|N}s_Q^T(q^{\rho + R^{-}}).
$$
The prefactor can be obtained by carefully regularizing the infinite product,
see appendix A.
Finally, since
\eqn\wschur{
W_{RQ} = s_R(q^\rho) s_Q(q^{\rho + R})
}
we reproduce the result of \refs{\AOSV} for the large $N$
limit of $Z({\cal C}, N,g_s)_{\cal R}=
S_{0,{\cal R}}(N,g_s)$, to all orders in perturbation theory:
\eqn\capp{
\eqalign{
S_{0,{\cal R}}(N,g_s) &= s(N,g_s)
(-1)^{|R^+| + |R^-|}
q^{-{1\over 2} N(|R^+|+|R^-|)} q^{-{1\over 2} (k_{R^+}+k_{R^-})}\cr
&\times\sum_{Q} (-)^{|Q|} W_{R^{+}Q}(q)q^{|Q|N} W_{Q^T R^{-}}(q)
}
}
The normalization constant $s(N,g_s)$ is independent of $R$.
Similarly, using \dtdn , we can compute the large $N$ limit of
$
Z({\cal P}, N,g_s)=  1/S_{0,{\cal R}}(N,g_s)
$
of the pant amplitude:
\eqn\pant{
S_{0,{\cal R}}(N,g_s)^{-1}
= s(g_s,N)^{-1}q^{{1\over 2} N(|R^+|+|R^-|)} q^{{1\over 2} (k_{R^+}+k_{R^-})}
\sum_{Q} {W_{R^{+}Q}\over W^2_{R^+ 0}}
q^{|Q|N}
{W_{Q R^{-}}\over W^2_{R^- 0}}.
}

As we saw earlier in this section,
the Yang-Mills amplitude corresponding to a D4 brane
wrapping a divisor ${\cal D}= {\cal L}_{p_1} \rightarrow {\Sigma}_{g}$
in $X$ is given in \divy\
\eqn\YMg{
Z^{YM}(N,\theta,g_s) =
z^{YM}(N,\theta,g_s)
\sum_{{\cal R}\in U(N)}\; S_{0{\cal R}}^{2-2G} \;q^{{p\over 2} C_2({\cal R})^{U(N)}} \;
e^{i \theta C_1({\cal R})^{U(N)}}
}
The normalization of the partition function $z^{YM}$ is
ambiguous in the D-brane theory.
In \AOSV , motivated by the black hole physics,
the normalization was chosen to be
$$
z^{YM}(g_s,N) = q^{{(p+2G-2)^2 \over 2p}\rho(N)^2}e^{{N \theta^2 \over 2 pg_s}}
$$
where $\rho(N)^2$ is the norm of the Weyl vector
$$
\rho(N)^2={1\over 12}(N^3-N).
$$
With the above choice of normalization and the results for
the pant and cap amplitudes, the
$perturbative$ part of the
large $N$ expansion of the Yang-Mills theory can be
written as a sum over chiral blocks:
\eqn\YMex{
Z^{YM}(N,\theta,g_s) \approx \sum_{l=-\infty}^{\infty}
\sum_{R_1,\ldots R_{|2G-2|}} Z_{R_1,\ldots R_{|2G-2|}}^{+}(t+p\, {l}\, g_s)
{\bar Z^{-}}_{R_1,\ldots,R_{|2G-2|}}({\bar t}-p\, {l}\, g_s),
}
%
where
\eqn\ztop{
Z_{R_1,\ldots R_{|2G-2|}}^{+}(t,g_s) =
z^{+}(g_s,t)e^{-{t(|R_1|+\ldots |R_{|2G-2|}|)\over p+2G-2}}
\sum_{R^+} {W_{R_1 R^+}\ldots W_{R_{|2G-2|}R^+}\over
W_{R^+ 0}(q)^{4g-4}}\;
q^{{p+2G-2\over 2}k_{R^+}} \; e^{-|R^+|t}
}
for $G \geq 1$ and
\eqn\ztopsphere{
Z_{R_1, R_{2}}^{+}(t,g_s) =
z^{+}(g_s,t)e^{-{t(|R_1|+|R_{2}|)\over p-2}}
\sum_{R^+}\;W_{R_1 R^+}\;W_{R^T_{2}R^{+}}\;
q^{{p-2\over 2}k_{R^+}} \; e^{-|R^+|t}
}
for $G=0$. In either case,
$$
t = {p+2G-2\over2} Ng_s - i \theta.
$$
The coefficient $z^{top}$ contains the
McMahon function $M(q)^{\chi(X)\over 2}$ which counts maps to points
in $X$, and
$$
z^{+}(g_s,t) = M(q)^{2-2G\over 2} e^{- {1\over p(p+2G-2)}{t^3 \over 6 g_s^2} + {p+2G-2 \over p}{t\over 24}}.
$$
As we will see in the next subsection, \ztop\ and \ztopsphere\
are precisely the topological string amplitudes on the geometry
under consideration.
For $G\neq 1$, these amplitudes contain contributions from
open strings ending on non-compact D-branes.
The integer $l$ appearing in the sum was interpreted in \torus\
as the RR 2-form flux through the Riemann surface $\Sigma$.

\subsec{Topological string amplitudes}

In this section we review the topological string amplitudes on ${\cal M}$,
$$
{\cal M} = {\cal L}_{p_1} \oplus {\cal L}_{p_2} \rightarrow {\Sigma_g}
$$
where
$$
p_1+p_2 = 2G-2
$$
Previously, we put $p_1 = -p$, $p_2 = p+2G-2$.
The closed topological string amplitudes on this manifold were solved
recently in \refs{\BP}. As explained there, the topological A-model on $X$
is effectively a topological field theory on $\Sigma$, which is completely specified by knowing the topological string amplitudes corresponding to
$\Sigma$ being a cap ${\cal C}$ and a pant ${\cal P}$, with specified degrees.

The cap amplitude is given by
$$
Z^{top}({\cal C};-1,0)_R = W_{R,0}(q) e^{-|R| t}
$$
where the numbers $(p_1,p_2)=(-1,0)$ denote the degrees of the line bundles,
and $t$ is the K\"{a}hler parameter of the cap.
This is nothing but the topological vertex\foot{The conventions of this paper
differ from \ref\AKMVtv{ M.~Aganagic, A.~Klemm, M.~Marino and C.~Vafa,
  ``The topological vertex,''
  Commun.\ Math.\ Phys.\  {\bf 254}, 425 (2005)
  [arXiv:hep-th/0305132].
} : $q$ in this paper is defined as $q=e^{-g_s}$ while
$q$ in \AKMVtv\ is $e^{g_s}$.
In particular, $C_{RQP}(q)$ of this paper
is defined to equal $C_{RQP}(q^{-1})$ of \AKMVtv .
When reading off the topological string amplitudes,
attention should be paid to this difference in conventions.}
of \AKMVtv\ with one stack of lagrangian branes:
$$
W_{R,0}(q) = C_{0,0,R}(q) q^{k_R/2}.
$$
The branes are wrapping the boundary of the disk on the Riemann
surface direction and extending in 2 directions of the fiber (see
\AKMVtv\ for a more detailed discussion of this).

The pant amplitude requires specifying three representations, corresponding to boundary conditions on the three punctures. It turns out that it vanishes, unless all three representations are equal, and it is given by
$$
Z^{top}({\cal P};1,0)_{R,R,R}
= {1\over W_{R,0}(q)} e^{-|R| t},
$$
where $t$ again measures the size of the pant.
The change of the bundle degrees is implemented by
the annulus operator
$$
Z^{top}({\cal A};-1,1)= q^{k_R/2}e^{-t|R|}
$$
These three amplitudes completely specify the closed string theory.
{}From these, any other amplitude can be obtained by gluing:
$$
Z((\Sigma_L \cup \Sigma_R);p^L_1+p^R_1,p^L_2+p^R_2)=\sum_Q\;
Z({\Sigma_L};p^{L}_1,p^{L}_2)_Q\;
Z(\Sigma_{R};p^{R}_1,p^{R}_2)_Q
$$
For example, the amplitude on a genus $g$ Riemann surface with bundle of degrees
$$
(p_1,p_2)=(-p,p+2G-2)
$$
the topological string amplitude is obtained by sewing $2G-2$ pant amplitudes
with $p+2G-2$ annuli operators ${\cal A}^{(-1,1)}$
giving
$$
Z({\Sigma_g};\;-p,\;p+2G-2) = \sum_R {W_{R,0}(q)^{2-2G}}
q^{{p+2G-2\over 2} k_R} e^{-|R| t}
$$

In addition to closed string amplitudes on ${\cal M}$, this allows us to
compute also open topological string amplitudes corresponding to
D-branes wrapping 1-cycles on $\Sigma$.  Moreover, we can also compute
topological string amplitudes with D-branes in the fibers over
$\Sigma$. As explained in \AOSV\ the results of \BP\ combined with
topological vertex results can be used to compute open topological
string amplitudes on ${\cal M}$ corresponding to with Lagrangian D-branes
which project to points on the Riemann surface.\foot{The crucial
property of ${\cal M}$ which \BP\ used in deriving their result was the
invariance under the torus action that rotates the fibers over
$\Sigma$. The Lagrangian branes of \AKMVtv\ preserve this symmetry,
which is why this generalization is possible.} We will need a slight
refinement of the statements there, so let us review the argument of
\AOSV\ in full.

The idea was to cut out a neighborhood in ${\cal M}$ of a point $P$ in
$\Sigma$ where the D-brane is. This is a copy of ${\bf C}^3$, and
topological string amplitudes with D-branes on ${\bf C}^3$ are solved by the
topological vertex. Gluing the ${\bf C}^3$ back in, we also glue the
topological string amplitudes on ${\bf C}^3$ and on $X-{\bf C}^3$ to get the
amplitudes on ${\cal M}$ with D-branes.

For example, consider a genus $G$
Riemann surface with a stack of D-branes over a point $P$.
Let the amplitude $before$ inserting the branes be
$$
Z(\Sigma; p_1,p_2) =\sum_{R} Z(\Sigma; p_1,p_2)_R
$$
and define an operator ${\cal O}_{QR}(P)$ which inserts the brane at
$P$ with boundary conditions given by $Q$, so that the amplitude with the D-brane is\foot{More precisely, fixing $Q$ is related to fixing a particular topological sector for worldsheet instantons
with boundaries on the brane. In string theory one naturally sums over these, so the quantity which appears is
$$
\sum_{Q} Z(\Sigma,P)_Q e^{-t_{C} |Q|} {\rm Tr}_{Q} U.
$$
Above, $t_{C}$ is the size of the holomorphic disk ending on the brane -- the opens string \Kahler modulus. Moreover $U$ is the holonomy of the D-brane gauge field.}

$$
Z(\Sigma,P; p_1,p_2)_Q =\sum_{R} {\cal O}(P)_{QR} Z(\Sigma; p_1,p_2)_R.
$$

We can compute
${\cal O}(P)_{QR}$ by cutting
out a neighborhood of $P$ out of ${\Sigma}$:
$$
Z(\Sigma; p_1,p_2) =\sum_{R}
Z({\cal C};\;-1,0)_R
Z(\Sigma-\{P\};\;p_1+1,p_2)_R
$$
and inserting it back, with D-branes.
To insert D-branes wrapping an $S^1$ in the ${\cal L}_{p_1}$ direction,
we replace
$$
Z({\cal C};\;-1,0)_R =C_{0,0,R}(q) q^{{k_R\over 2}}
$$
by
$$
Z({\cal C};\;-1,0)_{QR} = C_{Q^t,0,R}(q) q^{{k_R\over 2}} q^{-{k_Q\over 2}}
= W_{Q^t R^t}(q)
q^{k_R\over 2}.
$$
For D-branes along ${\cal L}_{p_2}$ direction, we get
$$
Z({\cal C};\;-1,0)_{QR} = C_{0,Q^t,R}(q) q^{{k_R\over 2}}= W_{Q R}(q).
$$
In writing the above, we have made a choice of framing for the D-brane,
where a different framing by $n$ units would amount to replacing
$
Z({\cal C})_{QR} \rightarrow Z({\cal C})_{QR} q^{-n k_Q/2},
$
and a choice of orientation of its world volume, which replaces
$
Z({\cal C})_{QR} \rightarrow Z({\cal C})_{Q^tR}.
$
We have made specific choices that will be convenient for us, but
for the purposes of our paper all the choices will turn out
equivalent, as we will see.

In summary, we found that operator inserting the branes along ${\cal L}_{p_1}$
direction is
$$
{\cal O}_{QR}(P)= {W_{Q^tR^t}\over W_{0R^t}}
$$
and the operator inserting a brane along ${\cal L}_{p_2}$ is
$$
{\cal O}_{QR}(P)={W_{QR}\over W_{0R}}
$$
where we used $W_{R0}(q) = q^{k_R\over 2} W_{R^t0}(q)$.

Using the above operators, we can compute the amplitudes corresponding to
inserting branes at arbitrary points $P_1,\ldots,P_k$ on ${\Sigma}$.
As explained in \AOSV\ the positions of the branes are complex structure
parameters, and correspondingly, the amplitudes are $indepentent$ of the
location of the points. More precisely, this is so as long as the points $P_i$
do not coincide, that is away from the boundaries of the moduli space of punctured Riemann surfaces.

When  $P_i$ and $P_j$ coincide, new contributions appear.
These are the amplitudes for a new holomorphic annulus and its multi-coverings,
connecting the two stacks of branes at $P_i$ and $P_j$,
without wrapping $\Sigma$.\foot{Even when $P_i$ and $P_j$ are
distinct points, there are holomorphic annuli connecting the two
stacks, but wrapping $\Sigma$, if both stacks of branes are in
the same fiber direction.}
If the two stacks are along the same fiber at a single base point,
the annulus amplitudes can be obtained from the one-stack case
by taking the holonomy matrix to be block-diagonal and then by
expanding in two unitary matrices.
If the two stacks lie on different fibers at a single base point,
we need to insert an operator constructed from
the full topological vertex $C_{Q_i Q_j R}$ depending on the boundary conditions
$Q_i$ and $Q_j$ on the two stacks of branes.
In this paper, we will not encounter these kinds of annulus amplitudes.

\subsec{Relation to the conjecture of OSV}

Combining the results of the previous two subsections, we now see that
counting of BPS states of N D4 branes with D2 and D0 branes in this geometry
precisely realize the OSV conjecture.
In particular, at large $N$, the D-brane partition function factorizes into
sum over chiral blocks, with chiral amplitudes \ztop\ and \ztopsphere\
which are
precisely the topological string amplitudes in this geometry \AOSV .
Namely,
$$
Z^{YM}(N,\theta,g_s) \approx \sum_{l=-\infty}^{\infty}
\sum_{R_1,\ldots R_{|2G-2|}}
Z_{R_1,\ldots R_{|2G-2|}}^{+}(t+p\, {l}\, g_s)
{\bar Z^{-}}_{R_1,\ldots,R_{|2G-2|}}({\bar t}-p\, {l}\, g_s),
$$
where
$$
Z_{R_1,\ldots R_{|2G-2|}}^{+}(t, g_s)=
Z_{R_1,\ldots R_{|2G-2|}}^{top}(t, g_s)
$$
corresponds to topological string amplitudes with insertions of
$|2G-2|$ ``ghost'' D-branes along the normal bundle to the divisor
${\cal D}$.
The \Kahler modulus $t$ associated to the
size of the base Riemann surface $\Sigma_G$ can be read off from \ztop\
and \ztopsphere\
to be
$$
t = {1\over 2} (p+2G-2) \; N g_s + i \theta.
$$
As explained in \AOSV\ this is exactly as predicted by the attractor mechanism and the OSV
conjecture. Namely, the attractor mechanism sets
$$
Re(t) = {1\over 2} {\#({\cal D} {\cap} \Sigma)} \;N g_s
$$
where the intersection number enters because the effective magnetic
charge of a D4 brane on ${\cal D}$ is set by the intersection number
of ${\cal D}$ with ${\Sigma}$, the two-cycle wrapped by the
D2-branes. In the present context, the intersection number can be
computed by deforming ${\cal D}$ along a section of its normal
bundle $O(p+2G-2)$ in ${\cal M}$. As this has $p+2G-2$ zeros, this
gives
$$
{\#({\cal D} {\cap} \Sigma)} = p+2G-2,
$$
exactly as expected.

Now consider the open string moduli associated to insertions
of ghost D-branes along the normal bundle.
These would enter the topological string amplitude
as
$$
Z^{{+}}_{R_1\ldots R_{|2G-2|}} \sim e^{-\sum_{i=1}^{|2G-2|}t_{C_i}|R_{i}|}
$$
where the $t_{C_i}$ is
the
complexified size of the holomorphic disk $C_i$ ending on the $i$-th stack of branes
(see footnote 4). From \ztop\ and \ztopsphere\ we can read off:
$$
Re(t_{C_i}) = {1\over 2}  N g_s,
$$
As was explained in \anv\ this value is in fact exactly what is
expected.  Namely, while the open string moduli are not supported by
compact 2-cycles, they still can feel the charge of the D4 brane.  The
attractor mechanism fixes their values according to the intersection
number of the open 2-cycle ending on the Lagrangian branes with the
divisor D wrapped by the D4 brane,
$$
Re(t_{C_i}) = {1\over 2}
{\# ({\cal D} {\cap} C_i)} N g_s.
$$
Since the $C_i$ lies simply along
the line bundle normal to ${\cal D}$, the intersection numbers are
canonical,
$$ {\# ({\cal D} {\cap} C_i)} = 1, $$
which is exactly what we see in \ztop\ and \ztopsphere .
More precisely, at
each puncture so for each $i$, we will
get as many open string moduli as there are D
branes there, in this case infinitely many. However, only the
``center'' of mass degree of freedom will get fixed by the D4 brane
charge, as the differences between the holomorphic disks ending on
various branes have zero intersection with ${\cal D}$.

As explained in \anv , the open string moduli have an interpretation
as non-normalizable $closed$ \Kahler moduli in the dual picture where
the ghost branes deform the geometry.
The duality in question is nothing
but the open-closed string duality, which can be understood very
precisely in the topological string context. The insertions
of ``ghost'' branes which correspond to the amplitudes above are dual
to certain non-normalizable deformations of the A-model geometry.
Because of this, the fact
that the D4 brane charges affected the open string moduli in the way
they did, is in fact forced on us!\foot{Note that, while in the topological string theory we can use either language, in
the physical superstring we do not have this freedom -- since there
are no Ramond-Ramond fluxes, the only available interpretation is the
closed string one.}
To make the fact that the normalizable and non-normalizable \Kahler moduli are at the same footing manifest, define
$$
Z^{top}(t,g_s, U_1\ldots U_{|2G-2|}) = \sum_{R_1,\ldots R_{|2G-2|}}
Z_{R_1,\ldots R_{|2G-2|}}\Tr_{R_1}U_1\ldots \Tr_{R_{|2G-2|}}U_{|2G-2|},
$$
where $U_i$ are the $U(\infty)$ valued holonomies on the ghost branes
and parameterize non-normalizable deformations of the geometry in the
closed string language.  Note that we can write \YMex\ as
\eqn\YMexx{
Z^{YM}(N,\theta,g_s) \approx
\sum_{l=-\infty}^{\infty}
\int d{U_1}\ldots dU_{|2G-2|}
Z^{top}(t+ lp g_s;U_1\ldots U_{|2G-2|})
{\bar Z^{top}}({\bar t}-lpg_s ;U^{\dagger}_1,\ldots U^{\dagger}_{|2G-2|}),
}
%
%
This has a very natural interpretation.
The above expresses the fact
that we are explicitly working with an ensemble where the
electric charges of the non-compact modes are set to zero!
The magnetic charges are also set to zero, apart from those induced by the
$N$ D4 branes, which we discussed above.\foot{The
more general charges that one can turn on
have a physical interpretation, as explained in \anv\
in terms of counting BPS states in $two$ dimensions.}

\newsec{Non-perturbative corrections and baby universes}

As reviewed in section two,
in the large $N$ limit, the D-brane partition function factorizes,
schematically as
$$
Z^{YM}(N,g_s) \approx Z^{top}(N,g_s){\bar Z}^{top}(N,g_s).
$$
As explained in \refs{\DGOV} the right hand side over-counts fermion
states which appear in $Z^{YM}$. The configurations which are
over-counted are those in the figure below.
\bigskip
\centerline{\epsfxsize 1.0truein\epsfbox{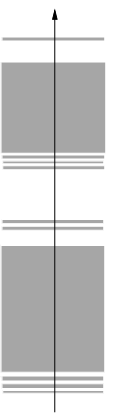}}
 \noindent{\ninepoint\sl
\baselineskip=2pt {\bf Fig.2}
{{Topological string amplitudes describe excitations about the top and the bottom fermi seas, as in the figure $1$, without restriction on the size of excitation.
Since $N$ is finite, this will over-count the configurations like those
in the figure. Here we cannot decide whether to include this in $Z^{top}$ or $\bar{Z}^{top}$,
and the configurations is thus counted twice in $|Z^{top}|^2$.
}}}
\bigskip
\noindent
This is a non-perturbative correction to the topological string amplitude.
There will be others, corresponding to a large number $N$ of fermions
splitting in arbitrary ways.

In this section we will first study the 2-black hole
correction in the large $N$ limit.
Then we will generalize the computations to the $M(\geq 3)$-black hole case.

\subsec{Non-perturbative corrections to cap and pant amplitudes}

As explained in \refs{\DGOV}, the fermion fluctuations around the
two fermi surfaces are independent only in perturbative $1/N$
expansion. Non-perturbatively, the two fermi surfaces are entangled.
The non-perturbative corrections to the factorization are described
by \DGOV\ splitting of $N$ fermions to groups with smaller numbers
of fermions. In the gravity theory, this corresponds to having
multiple black holes.  For example, consider the 2 black hole case:

\bigskip
\centerline{\epsfxsize 4.0truein\epsfbox{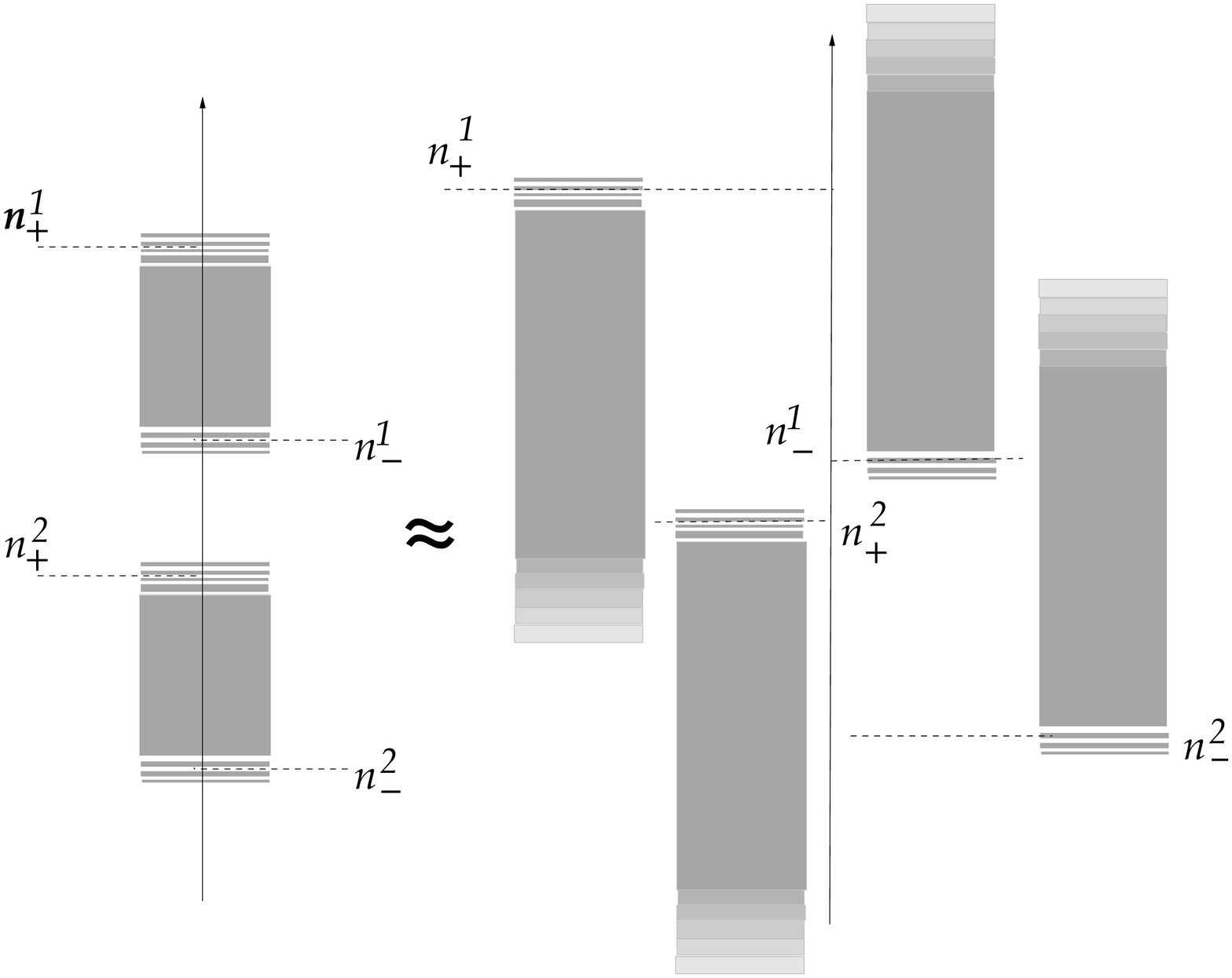}}
 \noindent{\ninepoint\sl
\baselineskip=2pt {\bf Fig.3} {{At large $N_{1,2}$ the configurations in Fig. 2 are described by
excitations of fermions about each of the 4 fermi sea surfaces. Each set of excitations
is described by a topological string partition function.
}}}
\bigskip

The $N$ fermions split into two groups of $N_1={n_1^+}-{n_1^-}$ and
$N_2={n_2^+}-{n_2^-}$ fermions where $n^{\pm}_i$ denote fermi-surface
momenta.  Following the same logic as before, we can immediately write down the
corresponding amplitude.  Let ${\cal R}$ and ${\cal Q}$ be
representations of $U(N_1)$ and $U(N_2)$ describing the
fermion states.
We can of course interpret ${\cal R} {\cal Q}$ as an irreducible representation of $U(N)$.
The amplitude we want to evaluate at large $N, N_1$ and $N_2$ is
$$
S_{0,{\cal R}{\cal Q}}(N,g_s).
$$
At large $N_{1,2}$, ${\cal R}$ and ${\cal Q}$ factorize as ${\cal R}_+{\bar {\cal R}_-}$
and ${\cal Q}_+{\bar {\cal Q}_-}$ with $U(1)$ charges determined by the locations of the fermi seas:
If the corresponding $SU(N)$ representations factorize as
${R_+{\bar R}_-}$ and $Q_+ {\bar Q}_-$ \refs{\moore,\AOSV}, then we define:
$$
{\cal R}_i^{+}: =  R_i^{+} + n_1^+,\quad {\cal R}_i^{-} := R_i^{-} - n_1^-,
$$
$$
{\cal Q}_i^{+} := Q_i^{+} + n_2^+,\quad {\cal Q}_i^{-} := Q_i^{-} - n_2^-,
$$
It is also useful to define the $U(1)$ charges of ${\cal R}$ and ${\cal Q}$:
$$
l_1:={n_1^+ + n_1^- \over 2}-{N_2\over 2},~~ l_2:={n_2^+ + n_2^- \over 2}+{N_1 \over 2}.
$$

Only the states near the fermi surfaces interact
in the amplitude we are computing, and the interaction depends only on
the distance between the fermions. The normalized amplitude thus
consists of self-interaction pieces of the two groups of fermions,
whose large $N$ limit we have just described:
$$
S_{0 ,{\cal R}}/S_{0,0}(N_1,g_s) \;S_{0, {\cal Q}}/{S_{0,0}}(N_2,g_s)
$$
and an interaction piece between them, consisting of
\eqn\one{
\prod_{i,j=1}^\infty
{[{R_i^+} -{Q_j^+}
+ n^+_1 - n^+_2 - i+j]\over
[n^+_1 - n^+_2 -i+j]}
{[-{R_i^-}+{Q_j^-} + n^-_1-n^-_2+i-j]\over
[ n^-_1-n^-_2+i-j]}
}
and
\eqn\two{
\prod_{i,j=1}^\infty
{[{R_i^+} + {Q_j^-} + n^+_1-n^-_2-i-j+1]\over
[n^+_1-n^-_2-i-j+1]}
{[-{R_i^-} - {Q_j^+} + n^-_1 - n^+_2 + i+j-1]\over
[ n^-_1 - n^+_2 + i+j-1]}
}
The contributions of fermi excitations around the bottom fermi seas at
$p=n_{1,2}^-$ come with a minus relative sign, since there fermion excitations have negative momenta relative to the fermi seas in their ground states.
While this derivation may appear heuristic, we will verify in the
appendix, via a careful calculation, that it is indeed correct.

It is
easy now to rewrite the large $N$ limit of the this in terms of
topological string objects. We will do so in a way which will be
convenient for us later.
We can write
$$
\eqalign{
S_{0, {\cal R}{\cal Q}}(g_s, N)&=
{\rm pr}_{{\cal R}{\cal Q}}\;
W_{0R^{+}}\,W_{0R^{-}}\,W_{0Q^{+}}\, W_{0Q^{-}}\; {\rm cr}_{{\cal R}{\cal Q}}.
}
$$
Here the prefactor ${\rm pr}$ is given by\foot{
We omit specifying the overall sign that will drop out when we take
the even power of the whole expression.}
\eqn\prdefone{\eqalign{
{\rm pr}_{{\cal R}{\cal Q}}&=\pm
M(q)^2 \eta(q)^{N}
q^{-{1\over 2} (k_{R^{+}}+k_{R^{-}}+k_{Q^{+}}+k_{Q^{-}})}
q^{-{N_1\over 2}(|R^{+}| + |R^{-}|)-{N_2\over 2}(|Q^{+}| + |Q^{-}|)}\cr
& \quad
\times
 q^{-{1\over 2} N_1 N_2 (l_1-l_2)}
q^{-{N_2\over 2}(|R^{+}| - |R^{-}|)+{N_1 \over 2}(|Q^{+}| - |Q^{-}|)},
}}
and the correction part ${\rm cr}$ by\foot{The infinite product, as it appears,
is not convergent and contains vanishing factors,
but it makes sense in terms of the finite products given in the
appendix.}
\eqn\crdefone{
\eqalign{
 {\rm cr}_{{\cal R}{\cal Q}}
&=\prod_{i,j=1}^\infty (1-q^{n_1^+ - n_1^- + R^{+}_i +R^{-}_j -i-j+1})
(1-q^{n_2^+ - n_2^- + Q^{+}_i +Q^{-}_j -i-j+1})\cr
&\times
(1-q^{n_1^+-n_2^+ + R^{+}_i -Q^{+}_j -i+j})
(1-q^{n_1^- - n_2^- - R^{-}_i +Q^{-}_j +i-j})\cr
&\times
(1-q^{n_1^+ - n_2^- + R^{+}_i +Q^{-}_j -i-j+1})
(1-q^{n_1^- - n_2^+ - R^{-}_i -Q^{+}_j +i+j-1}).
}}
${\rm cr}_{{\cal R}{\cal Q}}{}^{-1}$,
which enters the pant amplitude $S_{0, {\cal R}{\cal Q}}(g_s,
N)^{-1}$,
can be expanded by Schur functions with summations
over Young diagrams.
This is done in Appendix C.
As in the single black hole case, these Young diagrams represent interactions among different Fermi surfaces.
{}From the topological string point of view, the Young diagrams are ghost branes.
We only need the pant amplitude as we will focus on the $G\geq 1$ case,
though the corresponding expression for the cap amplitude is easy to obtain.
We will put off dealing with the correction part until
subsection 3.3.

\subsec{Leading block amplitudes in the two-universe case}

We now want to evaluate the two-universe contribution.
This can be obtained by evaluating the contribution to YM
amplitude of representations in Figure 2:
\eqn\YMtwo{
z^{YM}(N,\theta,g_s)
\sum_{{\cal R}, {\cal Q}}
S_{0,{\cal R}{\cal Q}}^{2-2G} (N,g_s)\;
q^{{p\over 2} C_2({\cal R}{\cal Q})^{U(N)}} \;
e^{i C_1({\cal R}{\cal Q})^{U(N)}\theta}
}

To simplify the algebra, let's first consider the leading chiral
block, where the interactions between the fermi seas are turned off. That
is, consider the piece of the amplitude where all ghost contributions
are set to zero. The pant amplitude in this
sector is
$$\eqalign{
S_{0, {\cal R}{\cal Q}}(g_s, N)^{-1}&=
{\rm pr}_{{\cal R}{\cal Q}}^{-1}\;
W_{0R^{+}}^{-1}W_{0R^{-}}^{-1}W_{0Q^{+}}^{-1}W_{0Q^{-}}^{-1},\cr
}$$
where ${\rm pr}_{{\cal R}{\cal Q}}$ is defined in \prdefone.
The Casimir $C_2({\cal R}{\cal Q})$ can be explicitly evaluated, see appendix
\casimirone .
We will show in the appendix C, that
\YMtwo\ can be expressed
in terms of topological string partition functions on ${\cal M}$:
\eqn\YMtwofin{\eqalign{
&\sum_{l'_1,l'_2=-\infty}^\infty
Z^{top}(t_1+p l'_1 g_s)\;
Z^{top}(t_2-p l'_1 g_s + (p+2G-2)l'_2g_s)\cr
&\times \bar{ Z}^{top}({\bar t}_2+p l'_1 g_s - (p+2G-2)l'_2 g_s)\;
\bar{ Z^{top}}(\bar{t}_1-p l'_1 g_s),
}}
where
$Z^{top}(t)$ is the closed topological string amplitude on ${\cal M}$ given in \ztop ,
corresponding to setting all the ghost representations to zero.
The K\"{a}hler moduli take the form
\eqn\kahler{
t_1 = {p+2G-2\over 2} N g_s+{p\over 2}k g_s - i \theta, \qquad t_2 = -{p\over 2}k g_s + i \theta.
}
Here $k=n_1^- - n_2^+$.\foot{The summed discrete parameters that shift the \Kahler moduli in \YMtwofin\ are defined by
$$
l'_1={n_1^++n_2^-\over 2},~~ l'_2={n_1^+-n_1^- - n_2^+ + n_2^- \over 2}.
$$}

What is the interpretation of this? To begin with, note that in addition to the divisor ${\cal D}$
that we wrapped D4 branes on,
$$
{\cal D} = {\cal L}_{-p}\rightarrow \Sigma
$$
there is another divisor ${\cal D}'$ in ${\cal M}$
$$
{\cal D}' = {\cal L}_{p+2G-2}\rightarrow \Sigma,
$$
that the branes could wrap.\foot{There are other divisors in ${\cal M}$, but only ${\cal D}$ and ${\cal D}'$ respect the toric symmetry of the Calabi-Yau, that we used heavily throughout.}

Now while
$$
\#({\cal D} \cap \Sigma)=p+2G-2
$$
we have
$$
\#({\cal D}' \cap \Sigma)=-p.
$$
This implies that \kahler\ is precisely consistent with attractor mechanism in the two universes
if we had
$$
(N,-k), \qquad (0,k)
$$
branes along $({\cal D}, {\cal D}')$
in the first and the second universes, respectively.

Note first of all that the net D4 brane charge is conserved: when the
baby universes split, we pair-create $k$ D4 branes and $k$ anti-D4
branes along ${\cal D}'$.  Moreover, even though in the first universe
we have both $N$ D4 branes on ${\cal D}$ and $-k$ D4 branes on ${\cal
D}'$, the effective D4 brane charge of either is positive.
Correspondingly, the two kinds of branes are mutually BPS, and the
first universe is supersymmetric.
However, by the same token, the
second universe preserves opposite supersymmetry: it is anti-BPS.
Since the baby universes are disconnected, the entire configuration
is still stable.

Note that in writing \kahler , we have flipped the sign of the theta-angle in the second universe. Correspondingly, relative to the BPS universe what we mean by the kets and the bra's get exchanged. The assignment is natural as CPT that exchanges
the BPS and the anti-BPS universes is an anti-linear transformation.
It is also natural in the context of the proof of the OSV
conjecture given in \proofone .
There factorization of $Z_{BH}$
was a consequence of
localization of membranes and anti-membranes to
the north and the south poles of the $S^2$ in a near horizon geometry of the black holes in M-theory compactification on Calabi-Yau ${\cal M}$.
The localization is due to supersymmetry: the supersymmetry preserved by either
 branes depends on their position on the $S^2$, and the
anti-membranes on one pole preserve the $same$ supersymmetry as
the membranes on the north pole.
In a BPS universe, the
topological string partition function came about from summing over
membrane states, and the anti-topological one over the anti-membrane states.
Going from a BPS to an anti-BPS universe these assignments should
naturally flip, as the membranes and anti-membranes get exchanged.
This exactly leads to the exchange of the kets and the bras we have
seen above.

\subsec{Ghost brane amplitudes in the two-universe case}

We now turn on ghost brane contributions and the subleading chiral
blocks. {}From \pantfin, we get that the full two-universe amplitude
can be written as follows:
\eqn\ghostamptwouniverse{
\eqalign{
\sum_{l'_1,l'_2=-\infty}^\infty \sum_{\{P\},\{P_1^\pm\},\{P_2^\pm\},\{S^\pm\}}&
\prod_{a=1}^{2G-2} N^{P_a}_{P_{1,a}^+P_{1,a}^-}
N^{P_a}_{P_{2,a}^+ P_{2,a}^-}\cr
\times & Z_1^{top} (t_1-p_1 l'_1 g_s)_{\{S^+\},\{P_1^+\}} \;
Z_2^{top} ({t}_2+p_1 l'_1 g_s + p_2 l'_2g_s)_{\{S^+\},\{P_1^-\}}\cr
\times&
{\bar Z}_2^{top}(\bar{t_2}-p_1 l'_1 g_s - p_2 l'_2 g_s)_{\{S^-\},\{ P_2^+ \}} \;
{\bar Z}_1^{top}(\bar{t}_1+p_1 l'_1 g_s)_{\{S^-\},\{P_2^-\}},
}}
where
$$
p_1=-p,\qquad p_2 = p+2G-2.
$$
Here, for example, $\{P_1^+\}$ denotes collectively a set of Young diagrams
$(P^+_{1,1},...,P^{+}_{1,|2G-2|})$. Explicitly,
\eqn\ghosttwoU{
\eqalign{
Z^{top}_1(t_1)_{\{S^+\},\{P_1^+\}}=&z^{\rm top}(g_s, t_1)
\sum_{R^+}e^{-|R^+|t_1}\, q^{{p+2G-2\over 2}k_{R^+}}\,
 W_{0R^+}^{2-2G}\,
\prod_a {W_{R^+ S^+_a}\over W_{0R^+}} {W_{R^+ P_{1,a}^+}\over W_{0 R^+}}
\cr
&~~\times e^{-g_s n_1^+|\{P_1^+\}|} e^{-{t_1\over p+2G-2}|\{S^+\}|},\cr
Z^{top}_2({t}_2)_{\{S^+\},\{P_1^-\}}=&z^{\rm top}(g_s, {t}_2)
\sum_{R^-}e^{-|R^-|{t}_2}\, q^{{p+2G-2\over 2}k_{R^-}}\,
W_{0R^-}^{2-2G}
\prod_a {W_{R^- S^+_a}\over W_{R^-}}
{ W_{{R^-}^T {P_{1,a}^-}^T}\over W_{{R^-}^T}}\cr
&~~\times e^{-g_s n_1^-|P_1^-|}(-1)^{|P_1^-|}e^{-{{t}_2\over p+2G-2}|S^+|},\cr
{\bar Z}^{top}_2(\bar t_2)_{\{S^-\},\{P_2^+\}}=&z^{\rm top}(g_s,\bar t_2)
\sum_{Q^+}e^{-|Q^+|{\bar t}_2}
 q^{{p+2G-2\over 2}k_{Q^+}}
W_{Q^+}^{2-2G}
\prod_a {W_{Q^+ S^-_a}\over W_{Q^+}} {W_{{Q^+}^T {P_{2,a}^+}^T}\over W_{{Q^+}^T}}\cr
&~~\times e^{g_s n_2^+|P_2^+|}(-1)^{|P_2^+|}e^{-{\bar t_2\over p+2G-2}|S^-|},\cr
{\bar Z}^{top}_1(\bar{t}_1)_{\{S^-\},\{P_2^-\}}=&z^{\rm top}(g_s, \bar{t}_1)
\sum_{Q^-} e^{-|Q^-|\bar{t}_1} q^{{p+2G-2\over 2}k_{Q^-}} W_{Q^-}^{2-2G}
\prod_a {W_{Q^- S^-_a}\over W_{Q^-}} {W_{Q^- P_{2,a}^-}\over W_{Q^-}}\cr
&~~\times e^{g_s n_2^-|P_2^-|}e^{-{\bar{t}_1\over p+2G-2}|S^-|},
}}
where $a$ runs from 1 to $2G-2$.
All of the $Z_i^{top}$, in the above are topological string amplitudes on ${\cal M}$, but
with different configuration of branes.
The two sets of ghost branes corresponding to representations labeled by
$\{S\}$ and $\{P\}$
that
appear, are qualitatively different.

The ghost branes associated to $\{P\}$ type representations correspond
to the kinds of the non-normalizable \Kahler moduli that
we had before.
The sum over $\{P\}$ type representations
can be efficiently replaced by an integral,\foot{
This uses the properties of the tensor product coefficients
$$
{\rm Tr}_{P_1} U  {\rm Tr}_{P_2} U = \sum_{P} N^{P}_{P_1P_2} {\rm Tr}_{P} U.
$$
}
and we can write \ghostamptwouniverse\ as
$$
\eqalign{
 \sum_{l'_1,\, l'_2=-\infty}^\infty \int \{dU\}
\sum_{\{S_+\}}&Z^{top}_1 (t_1-p_1 l'_1 g_s;\{U\})_{\{S_+\}}
Z^{top}_2 (t_2+p_1 l'_1 g_s + p_2 l'_2g_s;\{U\})_{\{S_+\}}\cr
\times
\sum_{\{S_-\}}
&{\bar Z}^{top}_2 (\bar{t}_2-p_1 l'_1 g_s - p_2 l'_2 g_s;\{U^{-1}\})_{\{S_-\}}
{\bar Z}^{top}_1 (\bar{t}_1+p_1 l'_1 g_s;\{U\})_{\{S_-\}}
\cr
},
$$
were $\{dU\}=\prod_{a=1}^{|2G-2|} dU_a$.
This is in precise agreement with expectations of
\DGOV : the above simply expresses
the fact that there are no net
electric charges turned on for the non-normalizable modes!
This had to be the case, since we have not turned them on in the theory on the D4 branes.

The only magnetic charges turned on for these modes are
those induced by the D4 branes.  To see this, we need to consider
the above amplitudes in more detail.
Consider first the anti-BPS universe. This has negative D4 brane charge
$(0,k)$ corresponding to $k$ D4 banes on ${\cal D}' =
{\cal L}_{p+2G-2}$.
The $\{P\}$ ghost branes in this universe are along the opposite line
bundle, i.e. along fibers of
$
{\cal L}_{-p}.
$
The intersection number of
${\cal D}'$ with the disks $C'$ ending on the ghost branes of that
universe are unambiguous and canonical
$$
\#({\cal D}' \cap C')=+1,
$$
and compute the effective
magnetic charge of the $k$ D4 branes:
This implies, as we discussed, that the sizes of the disks are fixed to
$$
{\rm Re}(t'_C)=
{1\over 2}k g_s.
$$
Since $(n_1^{-} - n_{2}^+)g_s =kg_s$,
this is in precise agreement with \ghostamptwouniverse .

In the universe with positive D4 brane charge $(N,-k)$
we have $N$ D4 branes along
${\cal D}$ and $-k$ along ${\cal D}'$. The
$\{P\}$ ghost branes lie along the fibers of
$
{\cal L}_{p+2G-2}
$
bundle. The intersections of disks along ${\cal D}'$ (ending on the $\{P\}$ ghost branes of that universe)
with ${\cal D}$ are unambiguous and equal to $1$, as we saw before. However, the intersections with ${\cal D}'$ are ambiguous. As
as $(n_1^{+} - n_{2}^-)g_s = (N+k)g_s$, the
result in \ghostamptwouniverse\
is consistent with
\eqn\fix{
\#({\cal D}' \cap C)=1,
}
implying
$$
{\rm Re}(t_C)=
{1\over 2}(N+k) g_s.
$$

The $\{S\}$ ghost branes are a new phenomenon, as they appear only when we have 2 or more universes. Before we turn to discussing them, let's consider the general pattern of the baby universe creation.

\subsec{Multi-universe amplitudes}
The analysis of the previous subsections can be straightforwardly generalized to the case where the parent universe splits into
$M$ baby universes.
%
The detailed computations are relegated to Appendix C.
The YM amplitude in the $M$-universe case is
\eqn\YMMone{\eqalign{
\sum_{\{S^{\pm}_{ij}\}}\int\{ dU\}
&\times Z^{top}(t_1;\{U\})_{\{S^{+}_{12}\},\{S^{+}_{13}\},...,\{S^{+}_{1M}\}}\cr
&\times  {Z}^{top}(t_{2},\{U\})_{
\{S^{+}_{12}\},\{S^{+}_{23}\},\{S^{+}_{24}\},...,\{S^{+}_{2,M}\}}\cr
&...\cr
&\times
Z^{top}({t}_{M}; \{U\})_{\{S^{+}_{1,M}\},\{S^{+}_{2,M}\},...,
\{S^{+}_{(M-1),M}\}} \cr
&\times  {\bar Z}^{top}(\bar t_{M},
\{U^{-1}\})_{\{S^{-}_{1,M}\},\{S^{-}_{2,M}\}, ...,\{S^{-}_{(M-1),M}\}}\cr
&...\cr
&\times {\bar Z}^{top}(\bar{t}_2,\{U^{-1}\})_{\{S^{-}_{1,2}\};\{ S^{-}_{2,3}\},...,
\{S^{-}_{2,M}\}}\cr
&\times \bar{Z}^{top}(\bar{t}_1;\{U^{-1}\})_{\{ S^{-}_{12}\}, \{S^{-}_{13}\},...,\{
S^{-}_{1,M}\}}
}
}
The \Kahler moduli are given by
$$\eqalign{
t_1=& {1\over 2}(p+2G-2) N g_s + {1\over 2} p k_1 g_s + i \theta,\cr
t_2= &- {1\over 2}(p+2G-2) k_1 g_s - {1\over 2} p k_2 g_s-i \theta,\cr
t_3=  &
{1\over 2}(p+2G-2) k_2 g_s + {1\over 2} p k_3 g_s+i\theta,\cr
\ldots\cr
t_{M} = & {1\over 2}(p+2G-2) k_{M-1} g_s + i\theta,
\; for\; M\;{odd}
}
$$
and
$$
t_{M} = - {1\over 2}p k_{M-1}g_s - i\theta,
\; for\; M\;{even}
$$
Here, we have omitted the discrete shifts in the \Kahler moduli for
simplicity. It is easy to see that this corresponds to the following
pattern of baby universe creation: Initially we have one BPS universe
with $(N,0)$ branes along $({\cal D}, {\cal D}')$ respectively.
Then, this splits into two universes, a BPS one with
$(N,-k_1)$ branes and
another anti-BPS with $(0,k_1)$ branes. In creating a new, third baby
universe, the universe with charge $(0,k_1)$ just created splits
into two, one with $(-k_2,k_1)$ branes which is anti-BPS, and another,
BPS one with $(k_2,0)$
branes, and so on.
$$
(N,0) \rightarrow (N,-k_1) \; \oplus\; (0,k_1)\;  \rightarrow (N,-k_1)
\oplus\; (-k_2,k_1) \; \oplus\; (k_2,0)
$$
It is easy to see that this pattern and attractor mechanism exactly reproduces
the \Kahler moduli above, with a sign of $\theta$ correlated with whether the universe is BPS or anti-BPS.
Moreover, throughout this process, the net electric and the magnetic charges for the non-normalizable \Kahler moduli stay turned off, apart from the magnetic charges induced by the D4 branes.
This explains \YMMone\ apart from the $\{S\}$ ghost branes which we turn to next.

\subsec{The $\{S\}$ ghost branes}

The $\{S\}$-ghost branes are pairwise interactions between the tops
(bottoms) of the Fermi seas, i.e. between the chiral (anti-chiral)
amplitudes.
Suppose we just created
a baby universe, so $M$ increased by one.  If the new baby universe is
BPS and effective D4 brane charge in it is of the form $(k,0)$,
the $\{S\}$ ghost branes in it are along the fibers of ${\cal
L}_{-p}$. Conversely, if it is anti-BPS with charge $(0,k)$,
the $\{S\}$ ghost branes are
along ${\cal L}_{p+2G-2}$.
Thus, the $\{S\}$-ghost branes are along
the line bundle where the D4 branes would have been in that universe.
Its interaction with the parent universes depends only on
whether they are BPS or anti-BPS. A parent BPS universe (one of the
$M$ original ones) has ghost branes along ${\cal L}_{-p}$ just like the
baby. If the parent is anti-BPS, the $\{S\}$ ghost branes are along
${\cal L}_{p+2G-2}$.

This suggests the following interpretation. Before we considered baby
universes, and in defining the partition function of the D4 branes, we
had a choice of the boundary conditions at infinity on
the D4 branes. In section two, we picked the bundle along the fibers to be flat, the only singularities of the bundle coming from D2 branes wrapping the Riemann surface and D0 branes bound to them.
We can in principle pick any choice of
boundary conditions at infinity that is consistent with
toric symmetry we used to compute the
partition function. In particular we can pick boundary conditions corresponding to having non-compact D2 branes at wrapping the fibers above $|2G-2|$ points on the Riemann surface.

Choosing
the boundary conditions means instead of computing
$$
\sum_{{\cal R}} S_{0{\cal R}}^{2-2G} q^{C_2({\cal R})} e^{i \theta C_1({\cal R})}
$$
we compute
\eqn\novel{
\sum_{{\cal R}} S_{{\cal Q}_1{\cal R}}\ldots S_{{\cal Q}_{|2G-2|}{\cal R}}/S_{0{\cal R}}^{|4 G-4|} q^{C_2({\cal R})} e^{i \theta C_1({\cal R})}.
}
This as corresponds to having D2 branes
wrapping the fibers at $|2G-2|$ points on the Riemann surface, where
${\cal Q}^{(a)}_i$, $a=1,...N$ of them bind to the $a$-th D4 brane.
This is because adding the D2 branes
with charges ${\cal Q}$ above a point $P$ on the Riemann
surface corresponds to inserting an operator
$$
{\rm Tr}_{\cal Q} e^{i \Phi(P)}
$$
in the path integral. This shifts the expectation value of the flux
$F$ through the disk $C_P$ centered at $P$
by
$$
\int_{C_P} F^{(a)} = {\cal Q}^{(a)}g_s, \qquad a=1,\ldots N
$$
since
$$
\int_{C_P} F^{(a)}= \oint_{P} A^{(a)},
$$
and $A$ and $\Phi$ are canonically
conjugate \AJS :
$$
\langle \oint_{P} A^{(a)} \;
{\rm Tr}_{\cal Q} \, e^{i \Phi(P)} \rangle = {\cal Q}^{(a)}g_s\; \langle {\rm Tr}_{\cal
Q}\,  e^{i \Phi(P)} \rangle.
$$
This has exactly the same effect as placing D-branes there of charges
${\cal Q}^a$, as claimed.
At large $N$, a $U(N)$
representation ${\cal Q}$ would factorize into two independent
representations $Q_+$ and $Q_-$, so this does not introduce any
additional correlations between the chiral and the anti-chiral
amplitude. However, the $Q_+$ and $Q_-$ dependence of this would look
like introducing ghost branes along the fibers wrapped by the D4
branes -- it would look precisely like the insertions of the $\{S\}$
ghost branes! From the closed string perspective, these ghost branes
also correspond to turning on non-normalizable deformations of the
closed string geometry, but these have their origin already in the
non-normalizable deformations of the D4 branes -- namely in turning on
charges of non-compact D2 branes.

The explanation of the $\{S\}$ ghost brane correlations is then as
follows.
The original BPS universe corresponds to $N$ branes along ${\cal D}$
with trivial boundary conditions at infinity, and no non-compact 2-brane charge.
When it fractionates as
$$
(N,0) \rightarrow (N,-k) \; \oplus\; (0,k).
$$
a baby anti-BPS universe is produced.
However, the 2 baby universes that result have non-compact D2 brane charges
turned on, but in such a way that the net brane charge is zero.
At the next step
$$
(N,0) \rightarrow (N,-k) \; \oplus\; (0,k)\;  \rightarrow (N,-k)
\oplus\; (-m,k) \; \oplus\; (m,0)
$$
a new BPS universe is created with $m$ units of D4 brane charge
along ${\cal L}_{p_1}$,
but also carrying some non-compact D2 brane charge.
Their charge is canceled by creating non-compact 2-branes
in the two older universes.\foot{There are two subtleties we should mention. If we took \novel\ on the
nose, the $\{S\}$-ghost branes and the $\{P\}$ ghost branes would be
inserted at the same points on the Riemann surface. In the large $N$
expansion of the amplitude at hand, the branes are not inserted at the
same points.
Moreover, from \YMMone\ it seems to follow that the
non-compact 2-brane charge is conserved pairwise, because the
correlations between the universes are pairwise. This appears rather
odd, because all other charges have only net charge conserved. The two
technical issues should be related. Namely, the reason the charges appear
pairwise conserved is that the ghost branes connecting one universe
with two others are not inserted at coincident points, but at
different points. Since they are inserted at different points it is
natural that the charges appear conserved only pairwise. In general, the amplitudes are independent of the location of the ghost-branes on the
Riemann surface branes, as long as these do not coincide.}

\vskip 1.0cm

\centerline{\bf Acknowledgments}

We would like to thank Nori Iizuka,
Marcos Marino, Natalia Saulina and Cumrun Vafa,
for valuable discussions.
The research of M.A. is supported in part by the UC Berkeley Center for
Theoretical Physics, a DOI OJI Award,
the Alfred P. Sloan Fellowship
and the NSF grant PHY-0457317.
The research of T.O. is supported in part by the NSF grants PHY-9907949 and PHY-0456556. H.O. is supported in part by
the DOE grant DE-FG03-92-ER40701 and the NSF grant OISE-0403366.

\appendix{A}{Infinite products as finite products}
In section two we explained how to take the large $N$ limit of our amplitudes, based on the free fermion description.
The free fermion description was easy to derive and convenient in terms of
making contact to topological strings.
Our derivation of the free fermion amplitudes was however heuristic.

Here we will show how to rewrite the infinite product
forms of the amplitudes -- as appear naturally when we think of
them as amplitudes of $N\rightarrow \infty$ of fermions, in terms
of finite products. The finite product form of the amplitudes is
easy to derive by direct algebraic manipulations (see Appendix B.)
We will begin by making contact with derivation in \AOSV\ for the one
universe case, as a warmup, and then proceed to more complicated ones.

\subsec{The case of one universe}

In section 3 we argued, based on the free fermion description
of the states that
\eqn\start{\eqalign{
S_{O{\cal R}}/S_{00}(N,g_s) &=
\prod_{1\leq i < j \leq \infty}
{[R^+_i - {R^+_j}+j-i]\over [j-i]}
\prod_{1\leq i < j \leq \infty}
{[{R^-_i} - {R^-_j}+j-i] \over [j-i]}\cr
&
\times \prod_{1\leq i, j \leq \infty}
{[{R^+_i} + {R^-_j}+  N -i-j+1]\over [N -i-j+1]}
}}
In this appendix we will explicitly regulate the infinite sums and show
that above is equal to the finite product form of the amplitude derived in
\AOSV .

Let's start with the first factor:
$$
\prod_{1\leq i < j \leq \infty} {[R^+_i - {R^+_j}+j-i]\over [j-i]}.
$$
Let $c_{R^+}$ denote the number of rows in $R^+$.
Then, we can write the above as
$$
\prod_{1\leq i<j\leq c_{R^+}}
{[R^+_i - {R^+_j}+j-i]\over [j-i]}.
\prod_{1\leq i\leq c_{R^+}; 1\leq j \leq \infty} {[R^+_i + c_{R^+}+j-i]
\over [{c_R^+}+ j-i]}.
$$
The infinite product factor can be written as a finite product
$$
\prod_{1\leq i\leq c_{R^+}; 1\leq j \leq \infty}
{[R^+_i + c_{R^+}+ j-i]\over [c_{R^+}+ j-i]}=
\prod_{1\leq i\leq c_{R^+}}
\prod_{1\leq \mu_i \leq R_i} {1\over [c_{R^+}+ \mu_i -i]}
$$
Putting the contributions together (see for example, \AKMVtv ),
$$
\eqalign{
\prod_{1\leq i < j \leq \infty} {[R^+_i - {R^+_j}+j-i]\over [j-i]}
&=
\prod_{1\leq i<j\leq c_{R^+}}
{[R^+_i - {R^+_j}+j-i]\over [j-i]}\prod_{1\leq i\leq c_{R^+}}
\prod_{1\leq \mu_i \leq R_i} {1\over [c_{R^+}+ \mu_i -i]}\cr
& = d_q(R^+).
}
$$
The answer above is the quantum dimension $d_q(R)$ of the symmetric group representation corresponding to $R$.
This can be rewritten in a perhaps more familiar form:
$$
d_q(R) = \prod_{\tableau{1}\in R} {1\over [h(\tableau{1})]},
$$
where $h(\tableau{1})$ is the hook length of the corresponding
box.

There are different ways to think
 about the answer above, however, as either the topological string amplitude
$$
d_{q}(R) = C_{R,0,0}q^{k_{R}/4} = W_{R,0}(q)
q^{k_{R}/4}
$$
or in terms of representation theory, as a Schur function.
$$
W_{R,0}(q) = s_R(q)
$$

Now, lets consider the last factor in \start :
$$
\prod_{1\leq i < j < \infty}
{[{R^+_i} + {R^-_j}+ N-i-j+1]\over [N -i-j+1]}
$$
We can again break it to finite products: denoting by $c_{R^-}$
the number of non-zero rows in $R^-$,
$$\eqalign{
\prod_{1\leq i ,j < \infty} {[{R^+_i} + {R^-_j}+ N-i-j+1]\over [N
-i-j+1]} &= \prod_{1\leq i < c_{R^+}; 1\leq j \leq c_{R^-}}
{[{R^+_i} + {R^-_j}+ N-i-j+1]\over [N-i-j+1]}\cr &\times
\prod_{1\leq i \leq {c_{R^+}};1\leq j \leq \infty} {[{R^+_i} +
N-c_{R^-}-i-j+1]\over [N-{c_{R^-}}-i-j+1]}\cr &\times \prod_{1\leq
i < \infty;1\leq j \leq {c_{R^-}}} {[{R^-_i} +
N-c_{R^+}-i-j+1]\over [N-{c_{R^+}}-i-j+1]}}
$$
The infinite products can be rewritten as follows:

$$
\eqalign{
 \prod_{1\leq i \leq {c_{R^+}};1\leq j < \infty}
{[{R^+_i} + N-c_{R^-}-i-j+1]\over [N-{c_{R^-}}-i-j+1]}& =
 \prod_{1\leq i \leq {c_{R^+}}; 1\leq j <\infty}
{[{R^+_i} + N-i-j+1]\over [N-i-j+1]}\cr
& \times
 \prod_{1\leq i\leq c_{R^+};1\leq j \leq {c_{R^-}}} {[N-i-j+1]\over [{R^+_i} + N-i-j+1]}\cr & =
 \prod_{1\leq i \leq {c_{R^+}}; 1\leq \mu_i \leq R_i}
{[N+\mu_i-i]}\cr
& \times
 \prod_{1\leq i \leq c_{R^+}; 1\leq j\leq c_{R^-}} {[N-i-j+1]\over [{R^+_i} + N-i-j+1]} }
$$
In the last step we rewrote the one infinite product as a finite product.

Now, we will put everything together.
First of all, recall that the quantum dimension $dim_q(R)$ of a representation $R$
of $U(N)$ can be written as
$$
dim_q(R)=\prod_{\tableau{1}\in R} {[N + j(\tableau{1}) - i(\tableau{1})]\over h(\tableau{1})}=
 \prod_{i= 1}^{c_{R^+}}\prod_{\mu_i=1}^{R_i}
{[N+\mu_i-i]} d_q(R).
$$
where $j$ counts the columns and $i$ the rows of the tableau.

Then, collecting all the factors, we have
$$
S_{O{\cal R}}/S_{00}(N,g_s) = dim_q(R_+) dim_q(R_-)
\prod_{i=1}^{c_{R^+}} \prod_{i=1}^{c_{R^-}} {[{R^+_i} + {R^-_j}+
N-i-j+1][N-i-j+1]\over [{R^+_i} + N-i-j+1][{R^-_i} + N-i-j+1] }
$$

This is the result in \AOSV .

\subsec{The case of two universes}

In a similar way to the one-universe case,
we will here rewrite the infinite products of
\one\ and \two\ as finite products.
In appendix B we will obtain them from
an honest computation involving only finite products.

Let us take the first factor in \one\
and write it as
$$\eqalign{
\prod_{1\leq i,j<\infty}
{[{R_i^+} -{Q_j^+}
+ n^+_1 - n^+_2 - i+j]\over
[n^+_1 - n^+_2 -i+j]}
&=\prod_{1\leq i\leq c_{R^+};1\leq j\leq c_{Q^+}}
{[{R_i^+} -{Q_j^+}
+ n^+_1 - n^+_2 - i+j]\over
[n^+_1 - n^+_2 -i+j]}\cr
&\times \prod_{1\leq i\leq c_{R^+};1\leq j<\infty}
{[{R_i^+}
+ n^+_1 - n^+_2+c_{Q^+} - i+j]\over
[n^+_1 - n^+_2+c_{Q^+} -i+j]}\cr
&\times \prod_{1\leq i < \infty;1\leq j \leq c_{Q^+}}
{[-Q^+_j
+ n^+_1 - n^+_2-c_{R^+} - i+j]\over
[n^+_1 - n^+_2 -c_{R^+}-i+j]}.
}
$$
The first infinite product on the r.h.s. can be rewritten as follows:
$$\eqalign{
 \prod_{1\leq i\leq c_{R^+};1\leq j<\infty}
{[{R_i^+}
+ n^+_1 - n^+_2+c_{Q^+} - i+j]\over
[n^+_1 - n^+_2+c_{Q^+} -i+j]}&=
\prod_{1\leq i\leq c_{R^+};1\leq j<\infty}
{[{R_i^+}
+ n^+_1 - n^+_2 - i+j]\over
[n^+_1 - n^+_2 -i+j]}\cr
&\times
 \prod_{1\leq i\leq c_{R^+};1\leq j\leq c_{Q^+}}
{[n^+_1 - n^+_2-i+j]\over
[{R_i^+}
+ n^+_1 - n^+_2- i+j]
}\cr
&=
\prod_{1\leq i\leq c_{R^+};1\leq\mu_i \leq R^+_i}
{1\over
[n^+_1 - n^+_2 -i+\mu_i]}\cr
&\times
 \prod_{1\leq i\leq c_{R^+};1\leq j\leq c_{Q^+}}
{[n^+_1 - n^+_2-i+j]\over
[{R_i^+}
+ n^+_1 - n^+_2- i+j]
}
}
$$
Other infinite products in \one\ and \two\ can be regularized
in the same way.
Combining everything, we find that the interaction piece between
two universes is
\eqn\apptwouniv{\eqalign{
&
\prod_{1\leq i\leq c_{R^+};1\leq j\leq c_{Q^+}}
{[{R_i^+} -{Q_j^+}
+ n^+_1 - n^+_2 - i+j][
 n^+_1 - n^+_2 - i+j]\over
[{R_i^+}
+ n^+_1 - n^+_2 - i+j][-{Q_j^+}
+ n^+_1 - n^+_2 - i+j]}\cr
&\times\prod_{1\leq i\leq c_{R^-};1\leq j\leq c_{Q^-}}
{[-{R_i^-}+{Q_j^-} + n^-_1-n^-_2+i-j][ n^-_1-n^-_2+i-j]\over
[-{R_i^-} + n^-_1-n^-_2+i-j][{Q_j^-} + n^-_1-n^-_2+i-j]}\cr
&\times\prod_{1\leq i\leq c_{R^+};1\leq j\leq c_{Q^-}}
{[{R_i^+} + {Q_j^-} + n^+_1-n^-_2-i-j+1][ n^+_1-n^-_2-i-j+1]
\over[{R_i^+} + n^+_1-n^-_2-i-j+1][{Q_j^-} + n^+_1-n^-_2-i-j+1]}\cr
&\times\prod_{1\leq i\leq c_{R^-};1\leq j\leq c_{Q^+}}
{[-{R_i^-} - {Q_j^+} + n^-_1 - n^+_2 + i+j-1][  n^-_1 - n^+_2 + i+j-1]
\over[-{R_i^-} + n^-_1 - n^+_2 + i+j-1][ -{Q_j^+} + n^-_1 - n^+_2 + i+j-1]}\cr
&\times\prod_{1\leq i\leq c_{R^+};1\leq\mu_i \leq R^+_i}
{[n_1^+-n_2^- -i +\mu_i]\over
[n^+_1 - n^+_2 -i+\mu_i]}\cdot\prod_{1\leq j\leq c_{Q^+};1\leq \mu_j \leq Q^+_j}
{[n_1^--n_2^+-\mu_j+j]\over[n_1^+-n_2^+-\mu_j+j]}\cr
&\times\prod_{1\leq i\leq c_{R^-};1\leq\mu_i \leq R^-_i}
{[n_1^- - n_2^+ +i-\mu_i]\over
[n_1^- - n^-_2 +i-\mu_i]}\cdot\prod_{1\leq j\leq c_{Q^-};1\leq \mu_j \leq Q^-_j}
{[n^+_1-n^-_2+\mu_j-j]\over[n_1^- - n^-_2 +\mu_j-j]}\cr
}}
\appendix{B}{Derivation in terms of finite products}

In this appendix, we will re-derive the interaction \apptwouniv\
between two universes
by an honest computation involving only finite products.
Let $M_1, M_2$ be integers such that
$c_{R^+}\leq M_1\leq N_1-c_{R^-}$,
$c_{Q^+}\leq M_2\leq N_1-c_{Q^-}$.
The final result will not depend on $M_1$ or $M_2$.
The interaction piece, up to the overall factor independent of
tableaux, is
$$\eqalign{
&\prod_{i=1}^{M_1}\prod_{j=1}^{M_2}{[ R^+_i-Q^+_j+n_1^+ - n_2^+  +j-i]
\over[n_1^+ - n_2^+  +j-i]}\cr
&\times\prod_{i=1}^{N_1-M_1}\prod_{j=1}^{N_2-M_2}
{[-R^-_i+Q^-_j+n_1^- - n_2^- - j+i]\over[n_1^- - n_2^- - j+i]}\cr
&\times\prod_{i=1}^{M_1}\prod_{j=1}^{N_2-M_2}
{[R^+_i+Q^-_j+n_1^+ - n_2^- -i-j+1]\over[n_1^+ - n_2^- -i-j+1]}\cr
&\times\prod_{i=1}^{N_1-M_1}\prod_{j=1}^{M_2}
{[ -R^-_i - Q^+_j+n_1^- - n_2^+ +i+j-1]
\over[n_1^- - n_2^+ +i+j-1]},\cr
}
$$
which can be rewritten as
\eqn\appone{\eqalign{
&\prod_{i=1}^{M_1}\prod_{j=1}^{M_2}{[ R^+_i-Q^+_j+n_1^+ - n_2^+  +j-i][n_1^+ - n_2^+  +j-i]
\over[ R^+_i+n_1^+ - n_2^+  +j-i][-Q^+_j+n_1^+ - n_2^+  +j-i]}\cr
&\times\prod_{i=1}^{N_1-M_1}\prod_{j=1}^{N_2-M_2}
{[-R^-_i+Q^-_j+n_1^- - n_2^- - j+i][n_1^- - n_2^- - j+i]
\over[-R^-_i+n_1^- - n_2^- - j+i][Q^-_j+n_1^- - n_2^- - j+i]}\cr
&\times\prod_{i=1}^{M_1}\prod_{j=1}^{N_2-M_2}
{[R^+_i+Q^-_j+n_1^+ - n_2^- -i-j+1][n_1^+ - n_2^- -i-j+1]
\over [R^+_i+n_1^+ - n_2^- -i-j+1][Q^-_j+n_1^+ - n_2^- -i-j+1]}\cr
&\times\prod_{i=1}^{N_1-M_1}\prod_{j=1}^{M_2}
{[ -R^-_i - Q^+_j+n_1^- - n_2^+ +i+j-1][n_1^- - n_2^+ +i+j-1]
\over
[ -R^-_i+n_1^- - n_2^+ +i+j-1][ - Q^+_j+n_1^- - n_2^+ +i+j-1]}\cr
}}
times
\eqn\apptwo{\eqalign{
&\prod_{i=1}^{M_1}\prod_{j=1}^{M_2}{[ R^+_i+n_1^+ - n_2^+  +j-i][-Q^+_j+n_1^+ - n_2^+  +j-i]
\over [n_1^+ - n_2^+  +j-i]^2}\cr
&\times\prod_{i=1}^{N_1-M_1}\prod_{j=1}^{N_2-M_2}
{[-R^-_i+n_1^- - n_2^- - j+i][Q^-_j+n_1^- - n_2^- - j+i]
\over [n_1^- - n_2^- - j+i]^2}\cr
&\times\prod_{i=1}^{M_1}\prod_{j=1}^{N_2-M_2}
{[R^+_i+n_1^+ - n_2^- -i-j+1][Q^-_j+n_1^+ - n_2^- -i-j+1]
\over [n_1^+ - n_2^- -i-j+1]^2}\cr
&\times\prod_{i=1}^{N_1-M_1}\prod_{j=1}^{M_2}
{[ -R^-_i+n_1^- - n_2^+ +i+j-1][ - Q^+_j+n_1^- - n_2^+ +i+j-1]
\over [n_1^- - n_2^+ +i+j-1]^2
}.\cr
}}
Due to cancelations in the ratios,
we can change the index ranges and rewrite \appone\
as the first four products in \apptwouniv.
\apptwo\ reduces to
$$
\eqalign{
&\prod_{i=1}^{N_1}\prod_{j=1}^{N_2}{[ R^+_i+n_1^+ - n_2^+  +j-i][-Q^+_j+n_1^+ - n_2^+  +j-i]
\over [n_1^+ - n_2^+  +j-i]^2}\cr
&\times{[-R^-_i+n_1^- - n_2^- - j+i][Q^-_j+n_1^- - n_2^- - j+i]
\over [n_1^- - n_2^- - j+i]^2}.\cr
}
$$
It is easy to check that these agree with the last four
products in \apptwouniv.

\appendix{C}{Expanding $S_{0,{\cal RQ}}^{-1}$ and $S_{0,{\cal R}_1...{\cal
R}_M}^{-1}$}

We first aim to rewrite the ghost brane correction part  ${\rm cr}_{{\cal R}{\cal Q}}{}^{-1}$
as defined in \crdefone~in a convenient form.
To do this recall the ``coproduct'' property of Schur functions:
$$
s_{R}(x,y) = \sum_{P,Q} N^{R}_{PQ}s_{P}(x) s_{Q}(y)
$$
where $N^{R}_{PQ}$ are tensor product coefficients.\foot{
When $x=(x_1,\ldots,x_{N_1})$ and $y=(y_1,\ldots y_{N_2})$
this describes branching rules of $U(N_1+N_2) \rightarrow U(N_1)\times U(N_2),$
as long as $R$ is a representation obtained by tensoring copies of fundamental representation of $U(N_1+N_2)$.}
Using this, we will write
$$
\eqalign{
&\prod_{i,j=1}^\infty
(1-q^{n_1^+-n_2^+ + R^{+}_i -Q^{+}_j -i+j})^{-1}
(1-q^{n_1^--n_2^- - R^{-}_i +Q^{-}_j +i-j})^{-1}\cr
&\times
(1-q^{n_1^+ - n_2^- + R^{+}_i +Q^{-}_j -i-j+1})^{-1}
(1-q^{n_1^- - n_2^+ - R^{-}_i -Q^{+}_j +i+j-1})^{-1}
}
$$
as
$$
\eqalign{
&\sum_{P}
s_{P}(q^{n_1^+} q^{R^+ +\rho}, q^{n_1^-}q^{-R^--\rho})\;
s_{P}(q^{-n_2^+} q^{-Q^+ -\rho},q^{-n_2^-}q^{Q^- +\rho})\cr
=& \sum_{P, P^{\pm}_{1},{P^{\pm}_2}}
N^{P}_{P^{+}_1 P^{-}_1}\;
s_{P^{+}_1}(q^{n_1^+} q^{R^+ +\rho})\;s_{P^{-}_1}(q^{n_1^-}q^{-R^--\rho})\cr
&\quad \times N^{P}_{P^{+}_2 P^{-}_2} \;
s_{P^{+}_2}(q^{-n_2^+} q^{-Q^+ -\rho})\;s_{P^{-}_2}(q^{-n_2^-}q^{Q^- +\rho}).
}
$$
We apply \wschur\ to rewrite the Schur functions in terms of the ratios of $W$'s.
The result of the
calculation is
\eqn\pantfin{
\eqalign{
{\rm cr}_{{\cal R}{\cal Q}}{}^{-1}
=\sum_{P,P_1^\pm, P_2^\pm}N^{P}_{P^{+}_1 P^{-}_1} N^{P}_{P^{+}_2 P^{-}_2}
&{W_{R^+ S^+} \over W_{0R^+}} {W_{R^+ P_1^+} \over W_{R^+}}
 {W_{R^- S^+} \over W_{0R^-}} {W_{{R^-}^T {P_1^-}^T} \over W_{{R^-}^T}}\cr
\times
&{W_{Q^+ S^-} \over W_{0Q^+}} {W_{{Q^+}^T {P_2^+}^T} \over W_{{Q^+}^T}}
 {W_{Q^- S^-} \over W_{0Q^-}} {W_{Q^- P_2^-} \over W_{Q^-}}\cr
\times (-1)^{|P_1^-|}  (-1)^{|P_2^+|}&\;
 q^{N_1|S^+| +N_2|S^-|+n_1^+ |P_1^+| + n_1^- |P_1^-| - n_2^+ |P_2^+|- n_2^-|P_2^-|}.
}}

We now compute \YMtwo.
The dependence of the quadratic Casimir on $(R^\pm, Q^\pm)$ can be read off
from its relation to the fermion energy $C_2({\cal RQ})=2 E({\cal R}{\cal Q})+{\rm const}$.
Also note that $C_2(R^\pm=Q^\pm=0)=N_1 l_1^2+N_2 l_2^2+ N_1N_2(l_1-l_2)$.
We then have
\eqn\casimirone{
\eqalign{
C_2({\cal R}{\cal Q})=&N_1 l_1^2+N_2 l_2^2+ N_1N_2(l_1-l_2) +k_{R^+}+k_{R^-}+ k_{Q^+}+k_{Q^-}\cr
&\quad + 2n_1^+|R^+|
-2n_1^-|R^-|+ 2n_2^+|Q^+| -2n_2^-|Q^-|.
}}
In the sector where the ghost branes are turned off,  the YM amplitude \YMtwo\ becomes:
\eqn\YMthree{
{\eqalign{
\sum_{l_1,l_2, R_{\pm},Q_{\pm}} & q^{{(p+2G-2)^2 \over 2p}\rho(N)^2}
\;e^{{N \theta^2 \over 2 pg_s}}\;
M(q)^{4-4g}\cr
&
\times  q^{{p\over 2} N_1 l_1^2 + {p\over 2} N_2 l_2^2}
\; q^{p l_1(|R_+|-|R_-|) + p l_2(|Q_+|-|Q_-|)}\;
 q^{{(p+2G-2)\over 2} N_1 N_2 (l_1-l_2)} \cr
& \times q^{{(p+2G-2) N_2\over 2}(|R_+| - |R_-|)}\;
q^{-{(p+2G-2)N_1\over 2}(|Q_+| - |Q_-|)}
e^{i\theta (N_1 l_1+|R_+|-|R_-|)}e^{i\theta (N_2 l_2+|Q_+|-|Q_-|)} \cr
&\times
\; q^{{(p+2G-2)\over 2} (k_{R_+}+ k_{R_-})}
\; q^{{(p+2G-2)\over 2} (k_{Q_+}+ k_{Q_-})}
\; q^{{(p+2G-2)N_1 \over 2}(|R_+| + |R_-|)}\;
q^{{(p+2G-2) N_2 \over 2}(|Q_+| + |Q_-|)}\cr
&\times
{W^{2-2G}_{ R^+}}\;
{W^{2-2G}_{R^-}}\;
{W^{2-2G}_{Q^+}}\;
{W^{2-2G}_{Q^{-}}}}}.
}
This expression can be rewritten in terms of closed topological
string partition functions as in \YMtwofin.
The YM amplitude with ghost brane contributions is given in
\ghosttwoU.

In the $M$-universe case, the pant amplitude is
$$\eqalign{
&S^{-1}_{0 {\cal R}_1...{\cal R}_M}(g_s, N)\cr
&=
\pm
M(q)^{-M}
\prod_{I=1}^M \eta(q)^{-N_I}
q^{{1\over 2} (k_{R_I^+}+k_{R_I^-})} q^{{N_I\over 2}(|R_I^+| - |R_I^-|)}
W_{R_I^+}^{-1}W_{R_I^-}^{-1}\cr
&\times \prod_{1\leq I<J\leq M} q^{{1\over 2} N_I N_J (l_I-l_J)}
q^{{N_J\over 2}(|R_I^+| - |R_I^-|)-{N_I \over 2}(|R_J^+| - |R_J^-|)}\cr
&\times
\prod_I \prod_{i,j=1}^\infty (1-q^{n_I^+ - n_I^- + R_{I,i}^+ +R_{I,j}^- -i-j+1})^{-1}
\cr
&\times\prod_{I<J}
\prod_{i,j=1}^\infty
(1-q^{n_I^+-n_J^+ + R_{I,i}^+ -R_{J,j}^+ -i+j})^{-1}
(1-q^{n_I^--n_J^- - R_{I,i}^- +R_{J,j}^- +i-j})^{-1}\cr
&\times
(1-q^{n_I^+ - n_J^- + R_{I,i}^+ +R_{J,j}^- -i-j+1})^{-1}
(1-q^{n_I^- - n_J^+ - R_{I,i}^- -R_{J,j}^+ +i+j-1})^{-1}
.
}$$
The infinite products in the last three lines can be written in terms of
the $W$-functions as in the two-universe case.
We can again use the co-product property of Schur functions,
by putting the $2M$ Fermi surfaces into two groups, the upper ones
(with fluctuations $R_1^+, R_1^-,..., R_{(M+1)/2}^-$ if $M$ is odd)
and the lower ones (with fluctuations $R_{(M+1)/2}^-,...,R_M^-, R_M^+$).
The quadratic Casimir in the $M$-universe case decomposes as
$$\eqalign{
&C_2({\cal R}_1...{\cal R}_M)\cr
=&\sum_{I=1}^M\left( l_I N_I(l_I-\sum_{J<I} N_J +\sum_{J>I}N_J)
+k_{R^+_I}+ k_{R^-_I} +2 n_I^+ |R^+_I| -2 n_I^-|R^-_I|
\right),
}
$$
where we have defined the $U(1)$ charge of ${\cal R}_I$ by
$$
l_I:={1\over 2}\left(\sum_{J<I} N_J + n_I^+ + n_I^- -\sum_{J>I} N_J\right).
$$
These ingredients are used to express the YM partition function in the $M$-universe case in terms of topological string partition functions, as in
\YMMone\ .

\footatend\vfill\supereject\immediate\closeout\rfile\writestoppt
\baselineskip=14pt\centerline{{\bf References}}\bigskip{\frenchspacing%
\parindent=20pt\escapechar=` \input refs.tmp\vfill\eject}\nonfrenchspacing
\end